\documentclass[12pt]{article}
\usepackage{amssymb,amsmath}
\usepackage{epsfig}
\usepackage{latexsym}
\usepackage{graphicx}
\usepackage{amsthm}
\usepackage{mathtools}
\makeatletter
\newdimen\itex@wd%
\newdimen\itex@dp%
\newdimen\itex@thd%
\def\itexspace#1#2#3{\itex@wd=#3em%
\itex@wd=0.1\itex@wd%
\itex@dp=#2ex%
\itex@dp=0.1\itex@dp%
\itex@thd=#1ex%
\itex@thd=0.1\itex@thd%
\advance\itex@thd\the\itex@dp%
\makebox[\the\itex@wd]{\rule[-\the\itex@dp]{0cm}{\the\itex@thd}}}
\makeatother

\makeatletter
\newif\if@sup
\newtoks\@sups
\def\append@sup#1{\edef\act{\noexpand\@sups={\the\@sups #1}}\act}%
\def\reset@sup{\@supfalse\@sups={}}%
\def\mk@scripts#1#2{\if #2/ \if@sup ^{\the\@sups}\fi \else%
  \ifx #1_ \if@sup ^{\the\@sups}\reset@sup \fi {}_{#2}%
  \else \append@sup#2 \@suptrue \fi%
  \expandafter\mk@scripts\fi}
\def\tensor#1#2{\reset@sup#1\mk@scripts#2_/}
\def\multiscripts#1#2#3{\reset@sup{}\mk@scripts#1_/#2%
  \reset@sup\mk@scripts#3_/}
\makeatother

\makeatletter
\newbox\slashbox \setbox\slashbox=\hbox{$/$}
\def\itex@pslash#1{\setbox\@tempboxa=\hbox{$#1$}
  \@tempdima=0.5\wd\slashbox \advance\@tempdima 0.5\wd\@tempboxa
  \copy\slashbox \kern-\@tempdima \box\@tempboxa}
\def\slash{\protect\itex@pslash}
\makeatother

\def\clap#1{\hbox to 0pt{\hss#1\hss}}

\let\oldroot\root
\def\root#1#2{\oldroot #1 \of{#2}}
\renewcommand{\sqrt}[2][]{\oldroot #1 \of{#2}}

\DeclareSymbolFont{symbolsC}{U}{txsyc}{m}{n}
\SetSymbolFont{symbolsC}{bold}{U}{txsyc}{bx}{n}
\DeclareFontSubstitution{U}{txsyc}{m}{n}

\DeclareSymbolFont{stmry}{U}{stmry}{m}{n}
\SetSymbolFont{stmry}{bold}{U}{stmry}{b}{n}

\makeatletter
\def\re@DeclareMathSymbol#1#2#3#4{%
    \let#1=\undefined
    \DeclareMathSymbol{#1}{#2}{#3}{#4}}
\re@DeclareMathSymbol{\neArrow}{\mathrel}{symbolsC}{116}
\re@DeclareMathSymbol{\neArr}{\mathrel}{symbolsC}{116}
\re@DeclareMathSymbol{\seArrow}{\mathrel}{symbolsC}{117}
\re@DeclareMathSymbol{\seArr}{\mathrel}{symbolsC}{117}
\re@DeclareMathSymbol{\nwArrow}{\mathrel}{symbolsC}{118}
\re@DeclareMathSymbol{\nwArr}{\mathrel}{symbolsC}{118}
\re@DeclareMathSymbol{\swArrow}{\mathrel}{symbolsC}{119}
\re@DeclareMathSymbol{\swArr}{\mathrel}{symbolsC}{119}
\re@DeclareMathSymbol{\nequiv}{\mathrel}{symbolsC}{46}
\re@DeclareMathSymbol{\Perp}{\mathrel}{symbolsC}{121}
\re@DeclareMathSymbol{\Vbar}{\mathrel}{symbolsC}{121}
\re@DeclareMathSymbol{\sslash}{\mathrel}{stmry}{12}
\re@DeclareMathSymbol{\invamp}{\mathrel}{symbolsC}{77}
\re@DeclareMathSymbol{\parr}{\mathrel}{symbolsC}{77}
\makeatother

\makeatletter
\DeclareRobustCommand\widecheck[1]{{\mathpalette\@widecheck{#1}}}
\def\@widecheck#1#2{%
    \setbox\z@\hbox{\m@th$#1#2$}%
    \setbox\tw@\hbox{\m@th$#1%
       \widehat{%
          \vrule\@width\z@\@height\ht\z@
          \vrule\@height\z@\@width\wd\z@}$}%
    \dp\tw@-\ht\z@
    \@tempdima\ht\z@ \advance\@tempdima2\ht\tw@ \divide\@tempdima\thr@@
    \setbox\tw@\hbox{%
       \raise\@tempdima\hbox{\scalebox{1}[-1]{\lower\@tempdima\box
\tw@}}}%
    {\ooalign{\box\tw@ \cr \box\z@}}}
\makeatother

\makeatletter
\def\udots{\mathinner{\mkern2mu\raise\p@\hbox{.}
\mkern2mu\raise4\p@\hbox{.}\mkern1mu
\raise7\p@\vbox{\kern7\p@\hbox{.}}\mkern1mu}}
\makeatother

\begin{document}  

\begin{titlepage}
\begin{flushright}
UTTG-17-11\\TCC-019-11
\end{flushright}

\bigskip
\bigskip
\bigskip
\bigskip
\centerline{\Large \bf On Three-Dimensional Mirror Symmetry}
\bigskip
\bigskip
\centerline{{\bf Anindya Dey}}

\bigskip
\centerline{Theory Group, Department of Physics and Texas Cosmology Center}
\centerline{The University of Texas at Austin,
TX 78712.}
\bigskip
\centerline{{ anindya@physics.utexas.edu} }
\bigskip
\bigskip

\begin{abstract}

\noindent Mirror Symmetry for a large class of  three dimensional $\mathcal{N}=4$ supersymmetric gauge theories has a natural explanation in terms of M-theory compactified on a product of $\text{ALE}$ spaces. A pair of such mirror duals can be described as two different deformations of the eleven-dimensional supergravity background $\mathcal{M}=\mathbb{R}^{2,1} \times \text{ALE}_{1} \times \text{ALE}_{2}$, to which they flow in the deep IR. Using the $A-D-E$ classification of $\text{ALE}$ spaces, we present a neat way to catalogue dual quiver gauge theories that arise in this fashion. In addition to the well-known examples studied in  \cite{Intriligator:1996ex}, \cite{deBoer:1996mp}, this procedure leads to new sets of dual theories. For a certain subset of dual theories which arise from the aforementioned M-theory background with an  $A$-type $\text{ALE}_{1}$ and a $D$-type $\text{ALE}_2$, we verify the duality explicitly by a computation of partition functions of the theories on $S^3$, using localization techniques . We derive the relevant mirror map and  discuss its agreement  with predictions from the Type IIB brane construction for these theories.

 \end{abstract}

\end{titlepage}


\tableofcontents






\section{Introduction}

Mirror symmetry is an extremely interesting example of duality in three-dimensional supersymmetric gauge theories with moduli spaces  (studied in \cite{Intriligator:1996ex},\cite{deBoer:1996mp} for $\mathcal{N}=4$ supersymmetry). The duality involves pairs of gauge theories  flowing to the same superconformal fixed point in the IR. In particular, mirror symmetry exchanges the Coulomb and the Higgs branches of the moduli spaces  of the two theories and leads to a linear map(known as the ``mirror map'') between the Fayet-Iliopoulos parameters associated with vector multiplets in one theory and masses of hypermultiplets in the other . The flavor symmetry at the superconformal fixed point is enhanced and this enhanced flavor symmetry group in the deep IR includes the flavor symmetry groups of the individual gauge theories as subgroups. We will restrict ourselves to theories with $\mathcal{N}=4$ supersymmetry in this work.\\

 A large class of three-dimensional supersymmetric gauge theories can be realized as world-volume theories on coincident D3 branes (with one compact direction) with appropriate boundary conditions at the two ends  \cite{Gaiotto:2008ak}, imposed by the presence of 5-branes (or orbifold/orientifold planes as we shall explain later) spanning directions orthogonal to the compact direction. Given a gauge theory with such a description, the mirror dual is obtained simply as the world-volume theory on the same set of coincident D3 branes with S-dual boundary conditions.  Therefore, mirror symmetry can be understood as a consequence of S-duality in Type IIB string theory.\\

In  \cite{Hanany:1996ie}, mirror duality between pairs of $A_{n-1}$ quiver gauge theories was deduced in this fashion and the construction was later extended to  $D_n$ quivers \cite{Kapustin:1998fa}.
In this note, we focus on the M-theory/ Type IIA realization of mirror symmetry for these quiver gauge theories, the basic idea for which was laid out in 
\cite{Porrati:1996xi}. We first discuss the complete M-theory description of the duality for pairs of $A_{n-1}$-type quivers and then show how the M-theory picture provides a nice way to catalogue mirror dual theories. In the process, we obtain new mirror pairs in addition to the more well-known examples discussed in \cite{Intriligator:1996ex},\cite{deBoer:1996mp} . We discuss the Type IIB realization for some of these theories. \\
The rest of the paper is devoted to studying the duality of mirror pairs using a direct comparison of their partition functions on $S^3$. The basic strategy is to show that the partition functions of the dual theories (each of which can be expressed as finite-dimensional integrals) are related by a simple redefinition of integration variables.
In addition to confirming the duality, this provides a nice way to read off the mirror map between dual theories. In \cite{Kapustin:2010xq}, such an analysis was performed for dual pairs involving $A_{n-1}$-type quivers. Here, we extend that computation to include certain theories involving $D_n$-type quivers and derive the mirror maps in such cases, comparing our results to predictions from Type IIB considerations.

The paper is organized as follows: Section 2 deals with the basic properties of $\mathcal{N}=4$ supersymmetric gauge theories in three dimensions and presents an elementary discussion on the Type IIB description of Mirror Symmetry. In section 3, we describe the M-theory picture and present a catalogue of mirror pairs that can arise in this fashion. Section 4 contains the partition function computation and derivation of the mirror maps.

\section{$D=3$, $\mathcal{N}=4$ Supersymmetric Gauge Theories}

\subsection{Massless Spectrum and R-symmetry}

$\mathcal{N}=4$ supersymmetry in $D=3$ has 8 real supercharges, which are doublets of $Spin(2,1)\sim SL(2,\mathbb{R})$ and transform as $(2,2)$ under the R-symmetry group, $SU(2)_R \times SU(2)_L$. The massless supermultiplets are most conveniently obtained by dimensional reduction of the $(1,0)$ supersymmetry in $D=6$, where the supercharges are four-component Majorana-Weyl spinors (in $D=6$) transforming as doublets of the $Sp(1)\sim SU(2)_{X}$ R-symmetry. The massless representations of the Poincare algebra are labelled by the representations of the little group $Spin(4)\sim SU(2)_1 \times SU(2)_2$ :$(2j_1+1,2j_2+1)$, where $j_1,j_2$ denote the ``{}spins''{} for the representations of $SU(2)_1$ and $SU(2)_2$ respectively. The particle content of the lowest massless representations (those without the graviton, or particles of yet-higher spin) of $D=6,\mathcal{N}=1$ can be summarized as follows:

Tensor multiplet:$(3,1;1)+\mathbf{(2,1;2)}+(1,1;1)$

Vector multiplet:$(2,2;1)+\mathbf{(1,2;2)}$

Half-hyper multiplet:$(1,1;2)+\mathbf{(2,1;1)}$

where we have indicated the representation of $SU(2)_1\times SU(2)_2\times SU(2)_X$, and denoted the fermions in bold.

On dimensional reduction to three dimensions, we embed $SU(2)_L$ diagonally in $SU(2)_1\times SU(2)_2$ and $SU(2)_R$ in $SU(2)_{X}$. The transformation properties of the dimensionally reduced fields (bosons and fermions respectively) under the R-symmetry group $SU(2)_L \times SU(2)_R$ can be summarized as:

Tensor multiplet:$(3\oplus1,1)+\mathbf{(2,2)}$

Vector multiplet:$(3\oplus1,1)+\mathbf{(2,2)}$

Half-hyper multiplet:$(1,2)+\mathbf{(2,1)}$

For the 3D vector multiplets (the reductions of 6D vector or tensor multiplets), the bosons consist of an $SU(2)_L$-triplet of scalars and a gauge boson, which is an R-symmetry singlet. Strictly speaking, the latter can be dualized to a circle-valued scalar only for an abelian gauge field. In that case, we obtain 4 scalars,transforming overall as $(3\oplus1,1)$. Nevertheless, for a non-abelian gauge field, the gauge symmetry (on either the Coulomb or Higgs branches) is higgsed to $\emph{at most}$ an abelian subgroup. So it is useful to carry over this counting in describing the low-energy theory.

The half-hypermultiplets transform in pseudo-real representation of the gauge group. Two copies of half-hypers give a $\mathcal{N}=4$ hypermultiplet in three dimensions.

Since the matter content in these supermultiplets is not symmetric with respect to the exchange $SU(2)_R \leftrightarrow SU(2)_L$, there can be ``twisted'' multiplets where $SU(2)_R$ and $SU(2)_L$ are exchanged.

\subsection{Lagrangian Description of $\mathcal{N}=4$ theories}

The action for the $\mathcal{N}=4$ supersymmetric gauge theories is most conveniently presented in the $\mathcal{N}=2$ superspace language. A $\mathcal{N}=4$ vector multiplet consists of one $\mathcal{N}=2$ vector multiplet and one $\mathcal{N}=2$ chiral multiplet in the adjoint of the gauge group. The action for a $\mathcal{N}=4$ quiver gauge theory consists of the following terms:

\begin{itemize}%
\item A Yang-Mills term for each factor in the gauge group.
\end{itemize}
\begin{equation}
S_{YM}=\frac{1}{g_{YM}^2}\int d^3x d^2\theta d^2\bar{\theta}(\frac
{1}{4} \Sigma^2-\Phi^{\dagger}\exp{2V} \Phi)
\end{equation}
where $\Sigma$ is a $\mathcal{N}=2$ linear multiplet, defined as $\Sigma=iD\bar{D}V$, where $V$ is a $\mathcal{N}=2$ vector multiplet, which is part of the $\mathcal{N}=4$ vector multiplet. $\Phi$ is a chiral multiplet in the adjoint of the gauge group.

\begin{itemize}%
\item Kinetic terms and minimal gauge couplings for the hypermultiplets.
\end{itemize}

\begin{equation}
S_{matter}=-\int  d^3x d^2\theta d^2\bar{\theta}\sum_i(\phi_i^{\dagger}\exp{2V} \phi_i+\tilde{\phi}_i^{\dagger}\exp{(-2V)} \tilde{\phi}_i)
\end{equation}

where $\phi_i,\tilde{\phi}_i$ are $\mathcal{N}=2$ chiral multiplets  constituting a $\mathcal{N}=4$ hypermultiplet (for each $i$) and the sum is over all the flavors in the theory.

\begin{itemize}%
\item Holomorphic superpotential term for the $\mathcal{N}=2$ chiral multiplets, compatible with the $\mathcal{N}=4$ supersymmetry.

\end{itemize}
\begin{equation}
S_{sup}=-i \sqrt{2} \int  d^3x d^2\theta \sum_i(\tilde
{\phi}_i\Phi\phi_i+c.c.)
\end{equation}

There are two possible deformations of the theory - namely, adding real or complex mass terms for the hypers or adding Fayet-Iliopoulos (FI) terms for the $U(1)$ factors in the gauge group.

\begin{itemize}%
\item The hypermultiplet masses transform as triplets of $SU(2)_L$ and can be interpreted as the lowest components of a background $\mathcal{N}=4$ vector multiplet coupled with the flavor symmetry currents in the usual way,
\end{itemize}

\begin{equation}
S_{mass}=-\int  d^3x d^2\theta d^2\bar{\theta}\sum_i(\phi_i^{\dagger}\exp{2V^i_m} \phi_i+\tilde{\phi}_i^{\dagger}\exp{(-2V^i_m)} \tilde{\phi}_i) -i\sqrt{2} \int  d^3x d^2\theta \sum_i(\tilde
{\phi}_i\Phi_m\phi_i+c.c.)
\end{equation}

where $V^i_m$ and $\Phi^i_m$ are respectively the $\mathcal{N}=2$ vector and chiral multiplets which make up the $\mathcal{N}=4$ background vector multiplet.
\begin{itemize}%
\item The FI factors transform as a triplet of $SU(2)_R$ and can be thought of as the lowest components of a \emph{twisted} $\mathcal{N}=4$ background vector multiplet, coupled to the topological currents for the $U(1)$ factors in the gauge group by a BF term,
\end{itemize}
\begin{equation}
S_{FI}=Tr\int  d^3x d^2\theta d^2\bar{\theta}\Sigma\hat{V}_{FI} +Tr(\int  d^3x d^2\theta \Phi\hat{\Phi}_{FI}+c.c.)
\end{equation}
where the $Tr$ picks up the $U(1)$ factors. The "  $ \hat{} $ " denotes a twisted multiplet.

\subsection{IR Behavior and Mirror Symmetry}

In the rest of the paper, we will be concerned with the infrared limit of the $\mathcal{N}=4$ supersymmetric gauge theories, where they approach a superconformal fixed point. The low-energy theory has a R-symmetry $Spin(4)\subset OSp(4|4)$ (the superconformal symmetry), which, in most cases, coincides with the manifest $SU(2)_R \times SU(2)_L$ in the Lagrangian description of the theory. For such theories, the conformal dimension for the bosonic fields is 1 for the vector multiplet and $\frac{1}{2}$ for the hypermultiplet. This obviously implies that the Yang-Mills term in the Lagrangian is irrelevant in the IR. The IR limit thus coincides with the strong coupling limit of the gauge theory, $g_{YM}\rightarrow \infty$.

The low-energy theory has a Coulomb branch where the expectation values of the hypermultiplet scalars are zero and the expectation values of the vector and adjoint scalars are in the Cartan subalgebra of the gauge group $G$. For a generic vev, the gauge group is broken to $U(1)^r$,where $r=rank(G)$. The Coulomb branch is a hyper-Kahler manifold of real dimension $4r$ , parametrized by vevs of one dual scalar and three adjoint scalars for every $U(1)$ factor. The Higgs branch, on the other hand, is characterized by non-zero vevs of the hypermultiplet scalars which break the gauge group completely. If the hypermultiplets transform in the representation $R$ of the gauge group $G$, then the Higgs branch is given by the hyper-Kahler quotient $M/G$ ($M$ being the quaternionic manifold parametrized by the scalar vevs of the hypermultiplet in the representation $R$ ) and parametrized by 4n real scalars, where $n=dim_{\mathbb{C}} (R)-dim_\mathbb {R} (G)$.

Two theories with $\mathcal{N}=4$ supersymmetry in $D=3$ (generically known as the A and the B model) are mirrors if they are related by an exchange of the Coulomb and the Higgs branch. This naturally implies an exchange of vector multiplets and hypermultiplets, $SU(2)_R$ and $SU(2)_L$ in the R-symmetry transformations and also the FI parameters and the hypermultiplet masses. Note that this 3D Mirror Symmetry is only a duality in the IR limit, and not at arbitrary energy scales.

In the next section, we discuss the Type IIB description of this duality for a discrete family of  $A_{n-1}$ quiver gauge theory pairs.

\subsection{Mirror Symmetry: Type IIB (Hanany-Witten) Interpretation}

The most commonly cited example of mirror duality involves the following discrete family of quiver gauge theory pairs:

A-model:$U(k)^m$ gauge theory with the matter content given by an extended $A_{m-1}$ quiver diagram (\ref{fig1l} (a)) with the $(m+1)$th node being identified with the first node. The $i$-th factor in the gauge group has $ w_i $ fundamental hypers, such that $\sum_i{w_i}=n$.

B-model:$U(k)^n$ gauge theory with the matter content given by an extended $A_{n-1}$ quiver diagram (\ref{fig1l} (b)) with the $(n+1)$th node being identified with the first node. The $j$-th factor in the gauge group has $v_j$ fundamental hypers, such that $\sum_j{v_j}=m$.

A simple counting of quaternionic dimensions of the Coulomb and Higgs branches shows, $dim M^A_C=dim M^B_H=m k$ and $dim M^A_H=dim M^B_C=n k$, as expected from duality. For a given sequence of integers $ \{w_i\}$  specifying the A-model, the sequence of integers $\{v_j\}$ in the B-model are determined by the following rule:

Consider a Young diagram with $m$ rows such that the length of the $p$-th row (the one at the top of the diagram is assigned $p=1$)is given by $\sum^{p}_{i=1}w_i$. Then the length of the $q$-th column of the Young diagram (the right-hand-most column is assigned $q=1$) is given as $\sum^{q}_{j=1}v_j$, thereby determining the set of integers $\{v_j\}$.

Finally, masses and FI parameters of the A-model are related to the FI parameters and masses of the B-model respectively by a set of linear equations, usually called ``mirror maps''. For the A-model, the total number of independent FI parameters is $m$ (one for each $U(1)$ in the gauge group) while the number of independent mass parameters is $n+m-m=n$ - accounting for the linear combination of masses that can be eliminated for every $U(1)$ factor in the gauge group by shifting the origin of the Coulomb branch. For the B-model, the total number of independent mass parameters is $m+n-n=m$ while the number of FI parameters is $n$, following the same reasoning as in the A-model. This gives a counting evidence of the fact that mirror duality exchanges masses and FI parameters. The precise form of the mirror map between FI parameters of the A-model and mass parameters of the B-model is given by \cite{deBoer:1996mp},

\begin{equation}
\sum^{i}_{l=1}\vec{\zeta}^{A}_l=\vec{m}^{B}_i
\end{equation}
or equivalently by,

\begin{equation}
\vec{\zeta}^{A}_l=\vec{m}^{B}_l -\vec{m}^{B}_{l-1}
\end{equation}
Mirror duality  for any pair of theories, described above, can be interpreted as a direct consequence of S-duality by resorting to a Type IIB brane construction involving D3, D5 and NS5 branes \cite{Hanany:1996ie} . Consider a Type IIB background, $\mathcal{M}=\mathbb{R}^{2,1}\times S^1 \times \mathbb{R}^3_{\vec{X}} \times \mathbb{R}^3_{\vec{Y}}$. A set of NS5 branes wrapping $\mathbb{R}^{2,1}\times \mathbb{R}^3_{\vec{X}}$ are located at different points in $S^1 \times \mathbb{R}^3_{\vec{Y}}$ while a set of D5 branes extend along $\mathbb{R}^{2,1}\times \mathbb{R}^3_{\vec{Y}}$ and are located at different points in the transverse space $S^1 \times \mathbb{R}^3_{\vec{X}}$. Finally, D3 branes wrapping $\mathbb{R}^{2,1}\times S^1$ can stretch between pairs of 5 branes located at adjacent points on the $S^1$.

The configuration of branes described above preserves eight real supercharges - appropriate amount of supersymmetry for $\mathcal{N}=4$ theories in three dimensions. Since the 5-branes are infinitely extended in two directions transverse to the 3-branes, the 5-brane world-volume fields have infinitely large volume coefficients in their kinetic terms compared to the world-volume fields on the D3 brane. As a result, 5 brane fields can essentially be treated as frozen backgrounds and enter the world-volume theory of the D3 branes as moduli in the effective QFT.

A D3 brane wraps the compact direction $S^1$, along which the (3+1)-dimensional world-volume theory can be Kaluza-Klein reduced to give a (2+1)-dimensional theory. Dynamical fields in this theory naturally arise from lightest excitations of open strings stretched between pairs of D-branes. Following the discussion in \cite{Hanany:1996ie}, the properties of this (2+1)-dimensional theory can be read off from the brane construction as follows:

$\bullet$ Open strings beginning and ending on $k$ coincident D3 branes stretched between two NS5 branes give a $\mathcal{N}=4,D=3$vector multiplet with a gauge group $U(k)$. The gauge coupling for the vector multiplet is given as,

\begin{equation}
\frac{1}{g^2_{YM}}=|s_1 - s_2|
\end{equation}
where $s_1,s_2$ are the positions of the two NS5 branes on $S^1$.

$\bullet$ Open strings beginning(ending) on $k$ coincident D3 branes stretched between the $i$th and the $(i+1)$th NS5 and ending(beginning) on $k'$ coincident D3 branes stretched between the $(i+1)$th and the $(i+2)$th NS5 give a $\mathcal{N}=4,D=3$ hypermultiplet in the bifundamental representation of $U(k) \times U(k')$.

$\bullet$ Open strings beginning(ending) on $k$ coincident D3 branes stretched between the $i$th and the $(i+1)$th NS5 and ending(beginning) on $p$ D5 branes whose $S^1$ coordinates lie between the two NS5 branes in question gives $p$ hypermultiplets in the fundamental representation of $U(k)$.

$\bullet$ The positions of the NS5 branes in $\mathbb{R}^3_{\vec{Y}}$ correspond to the FI parameters for the gauge groups. Assuming that all the D3 branes are located at the origin of $\mathbb{R}^3_{\vec{X}}$, the positions of the D5 branes in $\mathbb{R}^3_{\vec{X}}$ correspond to the masses of the fundamental hypers. The masses of the bifundamental hypers correspond to the relative position of D3 branes between the $i$th pair of NS5 branes with respect to the D3 branes between the $(i+1)$th pair along $\mathbb{R}^3_{\vec{X}}$, which we set to zero.

With these rules, one can immediately write down the appropriate Type IIB brane construction for the A-model. It simply consists of $m$ NS5 branes arranged in a circle with $w_i$ D5 branes located between the $i$th and the $(i+1)$th NS5 on the circle,such that $\sum_i w_i =n$. Finally, we have $m$ sets of $k$ D3 branes stretched between the $m$ pairs of NS5 branes, without any relative displacement along $\mathbb{R}^3_{\vec{X}}$.

Now,let us consider the action of S-duality on this system. NS5 and D5 branes transform into each other under S-duality while D3 branes remain invariant. The S-dual system consists of $n$ NS5 branes and $m$ D5 branes and one can easily see that the number of D5 branes between the $j$th and the $(j+1)$th NS5 is precisely $v_j$ (for all $j$) - the answer expected from the Young diagram rule stated above (after a possible cyclic permutation of brane labels).Therefore, the S-dual system clearly gives a Type IIB brane construction for the B-model: a $U(k)^n$ gauge theory described by a $A_{n-1}$ quiver and a distribution of $m$ fundamental hypers among its'{} $n$ nodes given by the sequence ${v_j}$. Also, since positions of NS5 and D5 are interchanged, the mapping of FI parameters to masses and vice-versa is trivially achieved.This is how mirror duality in three dimension can be understood as a consequence of S-duality in a Type IIB set-up.

In the next section, we present another description of the duality in terms of a Type IIA/M-theory. Obviously, the two descriptions are related via T-duality along $S^1$, but the M-theory picture, as we shall soon demonstrate, is helpful in classifying a large family of mirror duals and also finding new dual pairs.

\section{M-theory Perspective and Cataloging Mirror Duals}
\subsection{M-theory Description of Mirror Symmetry}
The M-theory description of mirror duality involves a solution of $11D$ supergravity with the following geometry, $\mathcal{M}: \mathbb{R}^{2,1}\times \text{ALE}_1 \times \text{ALE}_2$, where $\text{ALE}$ denotes an ``Asymptotically Locally Euclidean'' space  - a smooth hyper-Kahler resolution of an orbifold singularity of the form $\mathbb{C}^2/\Gamma$ ($\Gamma$ being a discrete subgroup of $SU(2)$). The \text{ALE} spaces have a well-known A-D-E classification, which is related to the A-D-E classification of the discrete subgroups of $SU(2)$ appearing in the orbifold limit of these spaces. In what follows, we will restrict ourselves to  $A$ and $D$ type \text{ALE} spaces, which can be further deformed to "Asymptotically Locally Flat"  or ALF spaces. These are four-dimensional hyper-Kahler manifolds, locally asymptotic to $\mathbb{R}^3 \times S^1$ at infinity (we will discuss a concrete case in a moment).\\
Consider first a simpler solution of $11D$ supergravity $\mathcal{M}=\mathbb{R}^{2,1}\times \mathbb{C}^2 \times \text{ALE}$, where the $\text{ALE}$ space has an $A_{n-1}$ singularity, i.e. this ALE space is the hyper-Kahler resolution of the orbifold singularity $\mathbb{C}^2/\mathbb{Z}_n$ .The $\text{ALE}$ metric is explicitly given as follows:

\begin{equation}
ds_{\text{ALE}}^2=H d\vec{r}^2 + H^{-1}(dx_{11}+\vec{C}.d\vec{r})^2
\end{equation}
where $H=\frac{1}{2}\sum_{i=1}^n \frac{1}{|\vec{r}-\vec{r_i}|}$ and $\nabla \times \vec{C}=-\nabla H$.

Now, we compactify the coordinate $x_{11}$ as $x_{11} \cong x_{11}+ 2 \pi g_{s} \sqrt{\alpha^{'}}$ and deform $H \to H^{'}$,where

\begin{equation}
H^{'}=\frac{1}{g_{YM}^2 \alpha^{'}}+ \frac{1}{2}\sum_{i=1}^n\frac{1}{|\vec{r}-\vec{r_i}|}   \label{deform}
\end{equation}
where $g_{YM}^2=g_s \frac{1}{\sqrt{\alpha^{'}}}$. This deformation gives a muti-centered Taub-Nut space, which is locally asymptotic to $\mathbb{R}^3\times S^1$ at infinity (and hence ``asymptotically locally flat''). At a generic point, the manifold is circle bundle over $\mathbb{R}^3$ and at the centers of the Taub-Nut $\vec{r}=\vec{r}_i$ the circle fiber shrinks to zero radius.

Note that the deformed background approaches the original background in the limit $g_{YM}\to \infty$.

This deformed M-theory background can be understood as the M-theory lift of a Type IIA supergravity background of $n$ $D6$ branes at transverse positions $\vec{r_i}$, wrapping the $\mathbb{C}^2$. In addition, if we have $k$ $M2$ branes wrapping the $\mathbb{R}^{2,1}$ in the M-theory picture, then these become $k$ $D2$ probe branes in the IIA picture. This Type IIA background preserves 8 real supercharges and as a result the world-volume gauge theory on the D2 branes has $\mathcal{N}=4$ supersymmetry. The gauge couplings are all proportional to $g_{YM}$ as defined above.

One can similarly deform an $\text{ALE}$ with a $D_{n}$-type singularity to an ALF form - the corresponding space turns out to be a generalization of the Atiyah-Hitchin space \cite{Sen:1997kz} . One can show that this space appears as the resolution of the orbifold singularity $\mathbb{C}^2/ \mathbb{D}_{n-2}$, where ${\mathbb{D}}_{n-2}$ denotes the Dihedral group with a $\mathbb{Z}_2$ central extension. The resultant Type IIA background consists of $n$ $D6$ branes at transverse positions $\vec{r_i}$ parallel to an $O6$ plane which wraps the $\mathbb{C}^2$.

Now, consider the manifold we are interested in, namely, $\mathcal{M}: \mathbb{R}^{2,1}\times \text{ALE}_1 \times \text{ALE}_2$, where the $\text{ALE}$ spaces are associated $A_{n-1}$ and $A_{m-1}$ singularities respectively. Deforming $\text{ALE}_1 \to \text{ALF}_1$, the resultant M-theory background $\mathcal{M}: \mathbb{R}^{2,1}\times \text{ALF}_1 \times \text{ALE}_2$ can easily be seen to correspond to a Type IIA background with $n D6$ branes wrapping $\text{ALE}_2$. The $k$ M2 branes become $k$ D2 branes in the Type IIA picture. The three-dimensional world-volume gauge theory given by the spectrum of $D2-D2$ and $D2-D6$ open strings is an $A_{m-1}$ quiver gauge theory with $n$ fundamental hypers associated with one of the $m$ $U(k)$ gauge groups. The gauge couplings are proportional to $g_{YM}$ which, in turn, is proportional to the radius of the circle fiber for $\text{ALF}_1$ at infinity (equation \ref{deform}).\\
One can similarly consider the deformed M-theory background $\mathcal{M}^{'}: \mathbb{R}^{2,1}\times \text{ALE}_1 \times \text{ALF}_2$, which leads to a Type IIA background with $m D6$ branes wrapping $\text{ALE}_1$. The three-dimensional world-volume gauge theory on the $D2$ branes, in this case, is an $A_{n-1}$ quiver gauge theory with $m$ fundamental hypers associated with one of the $n$ $U(k)$ gauge groups. The gauge couplings are proportional to $g^{'}_{YM}$ which is proportional to the radius of the circle fiber for $\text{ALF}_2$ at infinity.\\
In the IR limit, when $g_{YM},g'_{YM} \to \infty$, the two quiver gauge theories, in question, are described by the same M-theory background $\mathcal{M}: \mathbb{R}^{2,1}\times \text{ALE}_1 \times \text{ALE}_2$ and therefore have equivalent IR dynamics. The exchange of hypermultiplet masses and FI parameters can also be clearly seen. In the A-model (obtained by deforming $\text{ALE}_1$ to $\text{ALF}_1$), the parameters $\vec{r}^{(n)}_i$ appear as masses of the fundamental hypers while the parameters $\vec{r}^{(m)}_j$ are the FI parameters. In the B-model, their roles are reversed.\\

We are, however, not done yet since mirror duality, as argued in the previous subsection, exists between a discrete family of pairs of $A_{m-1}$ and $A_{n-1}$ quiver gauge theories, where each distinct pair can be characterized by a certain distribution $\{w_i\}$ of $n$ fundamental hypers on the $A_{m-1}$ quiver and a related distribution $\{v_j\}$ of $m$ fundamental hypers on the $A_{n-1}$ quiver. In the Type IIA picture, the  distribution of fundamental hypers can be achieved by turning on background fluxes for world-volume gauge-fields on D6 branes. For example, consider the A-model described previously. In the orbifold limit of the $\text{ALE}$ space which D6s are wrapping, the Wilson line $e^{i\int A_i}$ for the $i$-th D6 world-volume gauge field along a closed path around the orbifold singularity can have any value $z^{l_i}$, where $z=e^{2\pi i/m}$ and $l_i=0,1,...,m-1$. We now define the number of D6 branes with Wilson line $z^{l_i}$ as $w_i$, so that the $U(n)$ Wilson line diagonalizes into blocks of dimension $w_i$. This naturally leads to the required flavor-symmetry breaking $U(n) \to \prod_i U(w_i)$. An identical procedure for the B-model gives the flavor-symmetry breaking $U(m) \to \prod_j U(v_j)$, where $\{w_i\}$s and $\{v_j\}$s are related in a particular way described by the Young diagram rule of the previous section. Since the set of integers $\{w_i\}$ in the A-model essentially determines the set of integers $\{v_j\}$ in the B-model, the background fluxes in the two cases must be related to each other as well. This relation is best understood in the M-theory lift of these theories.

In the M-theory picture, the background gauge-field fluxes are lifted to appropriate 4-form G-fluxes in the purely geometrical background described so far. In the rest of the subsection, we write down an explicit expression for this G-flux in terms of a basis of self-dual two forms on the Taub-Nut introduced by Witten in \cite{Witten:2009xu} and show how it leads to the appropriate flavor-symmetry breaking in the mirror dual theories. For this purpose, we will consider a slightly more general M-theory background, $\mathbb{R}^{2,1}\times \text{ALF}_1 \times \text{ALF}_2$ (one can obviously take $\text{ALE}$ limits of the results whenever necessary).\\
Let us denote by ${\omega^{\alpha}_i}$ the set of anti-selfdual 2-forms on $\text{TN}_{\alpha}$ (Taub-Nut associated with a $A_{\alpha-1}$ singularity) which are curvature forms of $\alpha$ independent $U(1)$ gauge fields over $\text{TN}_{\alpha}$. Equivalently, these can be treated as curvature $(1,1)$ forms (in some chosen complex structure) over $\alpha$ complex line-bundles $\mathcal{L}_{i} \to \text{TN}_{\alpha}$, $i=1,...,\alpha$. In 
\cite{Witten:2009xu}, these 2-forms were shown to arise naturally in course of the hyper-Kahler quotient construction of a multi-centered Taub-Nut space and were explicitly constructed. On T-dualizing a Hanany-Witten system with $\alpha$ NS5 branes, $\omega^{\alpha}_i$s appear as the $\alpha$ B-field modes that are dual to the positions of the NS5s on the $S^1$. These 2-forms define a basis for $H^2(TN_{\alpha},\mathbb{Z})$ and obey the constraint that the two-form $\omega=\sum^{\alpha}_{i=1} \omega^{\alpha}_i$ has a vanishing integral over each compact cycle on the Taub-Nut.

The G-flux on the M-theory background $\mathbb{R}^{2,1}\times \text{ALF}_1 \times \text{ALF}_2$ can be described in a basis of the 4-forms ${\omega^{n}_i} \wedge {\omega^{m}_j}$ as follows:

\begin{equation}
G=\sum^{n}_{i=1}\sum^{m}_{j=1} a_{ij} {\omega^{n}_i} \wedge {\omega^{m}_j}    \label{gflux}
\end{equation}
where $a_{ij}$ is a $n \times m$ matrix with entries 0 or 1. For a given Young diagram representing the integers $\{w_i\}$ and $\{v_j\}$, $a_{ij}$ can be specified by the following rule:

$\bullet$  $a_{ij}=1$ if there'{}s a box at the $ij$-th position and is assigned the value 0 otherwise (the index $i$ of $\omega^{n}_i$ increases from bottom to top of the Young diagram while the index $j$ of $\omega^{m}_j$ increases from right to left).

The A-model corresponds to taking the circle direction of M-theory along $TN_{n}(\text{ALF}_1)$. On writing the G-flux as $G=\sum^{n}_{i=1} \omega^{n}_i \wedge F^{n}_i$, the $i$th D6 brane is endowed with the flux $F^{n}_{i}=\sum^{m}_{j=1} a_{ij} \omega^{m}_j$ (for $i=1,...,n$). For the B-model, the circle direction is chosen along $TN_{m}$ and writing $G=\sum^{m}_{j=1} F^{m}_j \wedge \omega^{m}_j$, we deduce that the flux on the $j$th D6 brane is $F^{m}_j=\sum^{n}_{i=1} a_{ij} \omega^{n}_i$ (for $j=1,...,m$). One can easily check that flux configurations constructed above lead to the appropriate flavor-symmetry breaking in the A and B-model respectively. Therefore, we obtain a complete M-theory description for mirror duality involving $A$-type quivers on both sides.

\begin{figure}[htbp]
\begin{center} 
\includegraphics[height=2.0in]{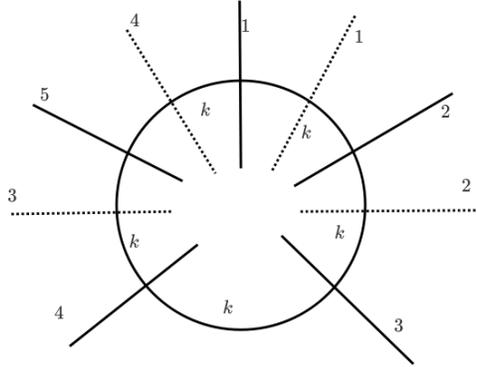}
\caption{Hanany-Witten set-up for a $U(k)^5$ quiver gauge theory with fundamental hypers $\{w_i\} =\{1,1,0,1,1\}$}
\label{fig0al}
\end{center}
\end{figure}

\begin{figure}[htbp]
\begin{center}
\includegraphics[height=2.0in]{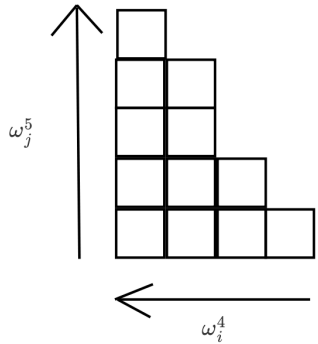}
\caption{Young Diagram encoding distribution of fundamental hypers for the $U(k)^5$ quiver gauge theory and its mirror dual }
\label{fig0bl}
\end{center}
\end{figure}

For a given $A_{n-1}$ quiver gauge theory, the background fluxes in the Type IIA picture can obviously be derived by T-dualizing the Type IIB brane construction \cite{Witten:2009xu}. Each D6 brane, obtained by T-dualizing a D5 brane along the circle direction, is associated with a Chan-Paton line bundle over $TN_{n}$. Suppose the $i$th D5 brane is located between the $\sigma_i$th and the $\sigma_{i+1}$th NS5 brane (with respect to some base point on the circle) in the compact direction. Then the Chan-Paton line bundle for the $i$th D6 brane is $\mathcal{R}_i=\otimes^{\sigma_i}_{k=1}\mathcal{L}^{-1}_{k}$. Therefore, the flux $F_i$ on the $i$th D6 brane (obtained on T-dualizing along the circle) is given as,

\begin{equation}
F_i=\sum^{\sigma_i}_{k=1} \omega^{n}_{k}  \label{Witten}
\end{equation}
One can easily check that the prescription for G-fluxes mentioned previously is consistent with the above equation (up to a possible relabeling of $\omega^{n}_k$s). We demonstrate this equivalence now using a toy example. Let the A-model be a $U(k)^5$ gauge theory with four fundamental hypers whose distribution on the quiver is given by the set $\{w_i\} =\{1,1,0,1,1\}$. The Type IIB brane construction for this theory is shown in figure \ref{fig0al} - the solid and dotted lines represent the NS5 branes and D5 branes respectively. Figure \ref{fig0bl} gives the Young diagram characterizing the distribution of fundamental hypers for this theory and its mirror dual. The mirror dual is therefore a $U(k)^4$ theory with five fundamental hyper distributed on the quiver as $\{v_j\}=\{1,1,2,1\}$.

In this case, the G-flux is given by the rule in equation \ref{gflux},

\begin{equation}
\begin{split}
G&=\omega^{4}_{1} \wedge \omega^{5}_{1}+\omega^{4}_{2} \wedge( \omega^{5}_{1}+\omega^{5}_{2})+
\omega^{4}_{3} \wedge( \omega^{5}_{1}+...+\omega^{5}_{4}) \\
&+ \omega^{4}_{4} \wedge(\omega^{5}_{1}+......+\omega^{5}_{5})
\end{split}
\end{equation}
On writing the G-flux as $G=\sum^{4}_{i=1} \omega^{4}_i \wedge F^{4}_i$, the fluxes can be read off from the above equation: $F^{4}_1=\omega^{5}_{1}, F^{4}_2=\omega^{5}_{1}+\omega^{5}_2, F^{4}_3=\omega^{5}_{1}+......+\omega^{5}_4, F^{4}_4=\omega^{5}_{1}+.........+\omega^{5}_5$, in obvious agreement with equation \ref{Witten}.  Note that the above configuration of fluxes also gives the required flavor symmetry breaking $U(4) \to U(1)^4$.

One can read off the fluxes in the B-model in a similar fashion by writing $G=\sum^{5}_{j=1} \omega^{5}_j \wedge F^{5}_j$ and show that they agree with equation \ref{Witten}, up to a relabeling of the $\omega^{4}_i$s. The flux configuration also leads to the correct flavor symmetry breaking $U(5) \to U(1)^3 \times U(2) $, in this case.

\subsection{Cataloging Mirror Duals}
The M-theory interpretation of mirror symmetry, described above, provides a neat way to catalogue a large class of mirror dual quiver gauge theories in terms of singularities appearing in the orbifold limit of the ALE spaces. A generic M-theory background $\mathcal{M}: \mathbb{R}^{2,1}\times \text{ALE}_1 \times \text{ALE}_2$ reduces to the orbifold $\mathbb{R}^{2,1}\times \mathbb{C}^2/\Gamma_1 \times \mathbb{C}^2/\Gamma_2$ in this limit, with the G-fluxes collapsing at the $\Gamma_1 \times \Gamma_2$ singularity. Below, we list the mirror duals that can be obtained for different choices of $\Gamma_1$ and $\Gamma_2$ (we will often refer to these as $\Gamma_1 \times \Gamma_2$ mirror dual theories), restricting ourselves to A and D-type singularities only.

In each case, we compute the quaternionic dimensions of the Coulomb and Higgs branches of the dual models and provide a counting evidence for the duality.

\begin{itemize}
\item \emph{$\Gamma_1=\mathbb{Z}_n$,$\Gamma_2=\mathbb{Z}_m$}

\end{itemize}

\begin{figure}[htbp]
\begin{center}
\includegraphics[height=3.0in]{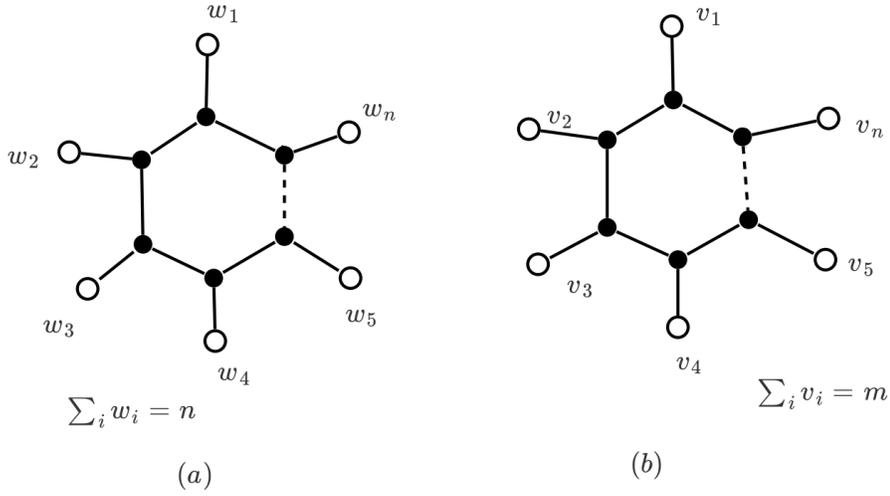}
\caption{Mirror Duals for $\Gamma_1=\mathbb{Z}_n$,$\Gamma_2=\mathbb{Z}_m$. Each black node denotes a $U(k)$ factor in the gauge group while the white node connected to it by a solid line denotes fundamental hyper(s) of that $U(k)$ }
\label{fig1l}
\end{center}
\end{figure}

A-model: $U(k)^m$ gauge theory with the matter content given by an extended $A_{m-1}$ quiver diagram (fig.\ref{fig1l} (a)) with the $(m+1)$th node being identified with the first node. The $i$-th factor in the gauge group has $w_i$ fundamental hypers, such that $\sum_i{w_i}=n$.

\begin{equation}
dim M^A_C=m k
\end{equation}
\begin{equation}
dim M^A_H=n k
\end{equation}
B-model:$U(k)^n$ gauge theory with the matter content given by an extended $A_{n-1}$ quiver diagram (fig.\ref{fig1l} (a)) with the $(n+1)$th node being identified with the first node. The $j$-th factor in the gauge group has $v_j$ fundamental hypers, such that $\sum_j{v_j}=m$.

\begin{equation}
dim M^B_C=n k
\end{equation}
\begin{equation}
dim M^B_H=m k
\end{equation}

\begin{itemize}%
\item \emph{$\Gamma_1=\mathbb{Z}_{n}$,$\Gamma_2=\text{Trivial}$}

\end{itemize}

\begin{figure}[htbp]
\begin{center}
\includegraphics[height=3.0in]{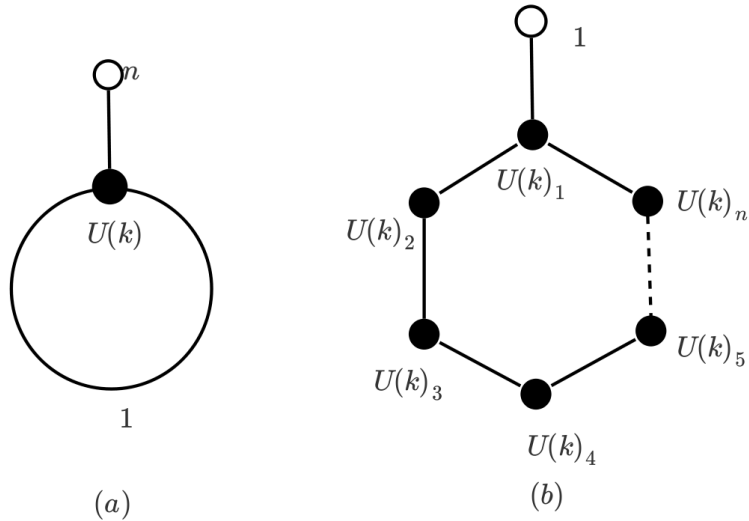}
\caption{Mirror Duals for $\Gamma_1=\mathbb{Z}_n$,$\Gamma_2=Trivial $.The fundamental hypers are denoted by white nodes while the adjoint hyper of the A-model is drawn as a circle beginning and ending on the black node, representing the $U(k)$ vector multiplet.
}
\label{fig2l}
\end{center}
\end{figure}

A-model: $U(k)$ gauge theory with n fundamental hypers and one hyper in the adjoint of $U(k)$ (fig. \ref{fig2l} (a)). 
\begin{equation}
dim M^A_C=k
\end{equation}
\begin{equation}
dim M^A_H=n k
\end{equation}
B-model: $U(k)^n$ gauge theory with the matter content given by an extended $A_{n-1}$ quiver diagram (fig. \ref{fig2l} (b)) with the $(n+1)$th node being identified with the first node. One of the $U(k)$ factor which we label as $U(k)_1$ has one hyper in the fundamental representation.

\begin{equation}
dim M^B_C=n k
\end{equation}
\begin{equation}
dim M^B_H=k
\end{equation}

\begin{itemize}%
\item \emph{$\Gamma_1=\mathbb{D}_{n-2}$,$\Gamma_2=\text{Trivial}$}

\end{itemize}

\begin{figure}[htbp]
\begin{center}
\includegraphics[height=3.0in]{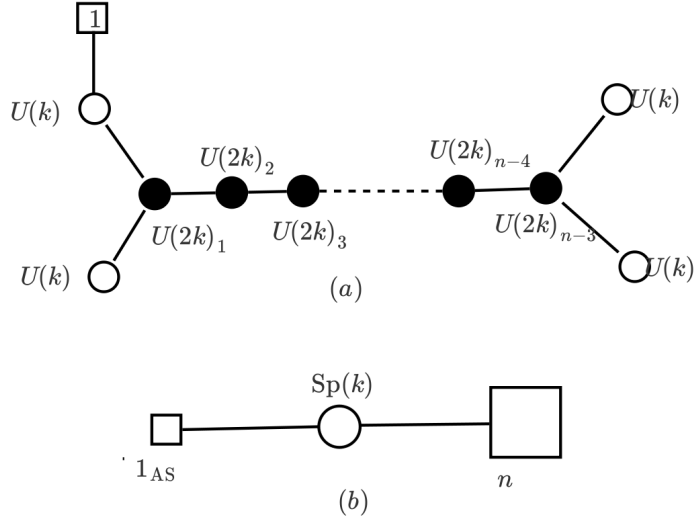}
\caption{Mirror Duals for $\Gamma_1=\mathbb{D}_{n-2}$,$\Gamma_2=\text{Trivial}$. In figure (a), the black nodes denote the $U(2k)$ factors, the white ones denote the $U(k)$ factors and the single fundamental hyper is drawn as a white square connected to the appropriate $U(k)$ factor by a solid line. In figure (b), the white node denotes the vector multiplet, while the large and the small square represent the set of $n$ fundamental hypers and the single antisymmetric hyper respectively.}
\label{fig3l}
\end{center}
\end{figure}

A-model:$U(k)^4 \times U(2k)^{n-3}$ gauge theory with the matter content given by an extended $\mathbb{D}_{n-2}$ quiver diagram (fig. \ref{fig3l} (a)). One of the $U(k)$ factors has one fundamental hyper. 

\begin{equation}
dim M^A_C=2 k (n-1)
\end{equation}
\begin{equation}
dim M^A_H=k
\end{equation}

B-model: $Sp(k)$ gauge theory with $n$ fundamental hypers and one hyper in the anti-symmetric representation of $Sp(k)$ (fig. \ref{fig3l} (b)). 

\begin{equation}
dim M^B_C=k
\end{equation}
\begin{equation}
dim M^B_H=2 k (n-1)
\end{equation}

\begin{itemize}%
\item \emph{$\Gamma_1=\mathbb{Z}_n$,$\Gamma_2=\mathbb{D}_{m-2}$}

\end{itemize}
$\bullet$ \emph{n even (n{\tt \symbol{62}}2)}

\begin{figure}[htbp]
\begin{center}
\includegraphics[height=3.0in]{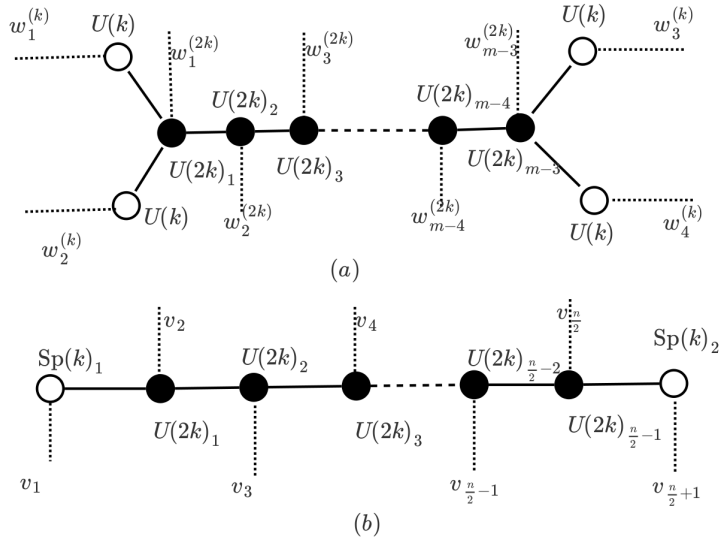}
\caption{Mirror Duals for $\Gamma_1=\mathbb{Z}_n$,$\Gamma_2=\mathbb{D}_{m-2}$, with $n$ even. Fundamental hypers are simply denoted as dotted lines connected to the appropriate black/white node.}
\label{fig4l}
\end{center}
\end{figure}

A-model:$U(k)^2 \times U(2k)^{m-3} \times U(k)^2$ gauge theory with the matter content given by a $\mathbb{D}_{m-2}$ quiver diagram (fig. \ref{fig4l} (a)). The $i$th factor in the gauge group has $w_i$ fundamental hypers, such that $\sum_i{w_i l_i}=n$, where $l_i=1,2$ is the Dynkin label for the $i$th node. 

\begin{equation}
dim M^A_C=2 k (m-1)
\end{equation}
\begin{equation}
dim M^A_H=n k
\end{equation}
B-model:$Sp(k)\times U(2k)^{\frac{n}{2}-1}\times Sp(k)$ gauge theory with bi-fundamentals (fig. \ref{fig4l} (b)). The $i$th factor in the gauge group has $v_i$ fundamentals, where $\sum_i{v_i}=m$.

\begin{equation}
dim M^B_C=n k
\end{equation}
\begin{equation}
dim M^B_H=2 k (m-1)
\end{equation}

$\bullet$ \emph{n odd (n{\tt \symbol{62}}1)}

\begin{figure}[htbp]
\begin{center}
\includegraphics[height=3.0in]{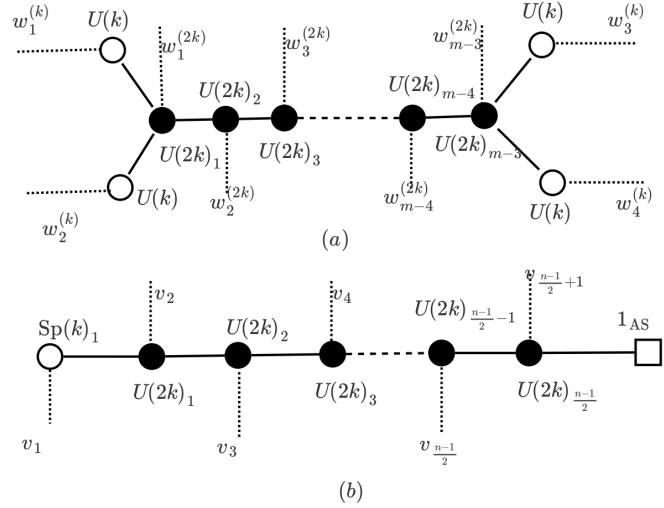}
\caption{Mirror Duals for $\Gamma_1=\mathbb{Z}_n$,$\Gamma_2=\mathbb{D}_{m-2}$, with $n$ odd}
\label{fig5l}
\end{center}
\end{figure}

A-model:$U(k)^2 \times U(2k)^{m-3} \times U(k)^2$ gauge theory with the matter content given by a $\mathbb{D}_{m-2}$ quiver diagram (fig. \ref{fig5l} (a)). The $i$th factor in the gauge group has $w_i$ fundamental hypers, such that $\sum_i{w_i l_i}=n$,where $l_i=1,2$ is the Dynkin label for the $i$th node.

\begin{equation}
dim M^A_C=2 k (m-1)
\end{equation}
\begin{equation}
dim M^A_H=n k
\end{equation}
B-model:$Sp(k)\times U(2k)^{\frac{n-1}{2}}$ gauge theory with bi-fundamentals and one hyper in the antisymmetric representation of $U(2k)_{\frac{n-1}{2}}$(fig. \ref{fig4l} (b)). The $i$th factor in the gauge group has $v_i$ fundamentals, where $\sum_i{v_i}=m$.

\begin{equation}
dim M^B_C=n k
\end{equation}
\begin{equation}
dim M^B_H=2 k (m-1)
\end{equation}

\begin{itemize}%
\item \emph{$\Gamma_1=\mathbb{D}_{n-2}$,$\Gamma_2=\mathbb{D}_{m-2}$}

\end{itemize}
$\bullet$ \emph{n,m even (n,m {\tt \symbol{62}} 4)}

\begin{figure}[htbp]
\begin{center}
\includegraphics[height=3.0in]{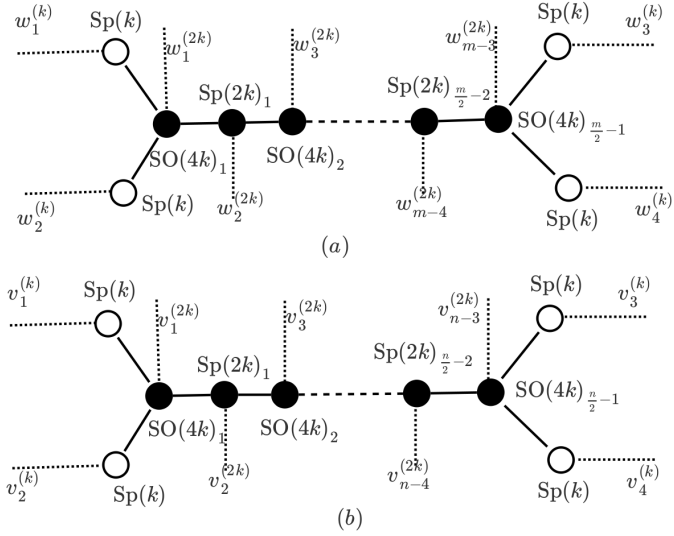}
\caption{Mirror Duals for $\Gamma_1=\mathbb{D}_{n-2}$,$\Gamma_2=\mathbb{D}_{m-2}$, with $n,m$ even.The solid lines connecting a pair of black nodes or a black and white node denote half-hypers in the bi-fundamental of the corresponding gauge groups. Fundamental hypers are denoted by dotted lines.}
\label{fig6l}
\end{center}
\end{figure}

A-model:$Sp(k)^2\times SO(4k)^{\frac{m}{2}-1}\times Sp(2k)^{\frac{m}{2}-2}\times Sp(k)^2$ gauge theory, with bi-fundamental half-hypers (fig. \ref{fig6l} (a)), described by a $\mathbb{D}_{m-2}$ quiver. The $i$th factor of the gauge group has $w_i$ fundamentals, such that $\sum_i{w_i}l_i=n$, where $l_i=1,2$ is the Dynkin label for the $i$th node. 

\begin{equation}
dim M^A_C=2 k (m-1)
\end{equation}
\begin{equation}
dim M^A_H=2 k (n-1)
\end{equation}
B-model:$Sp(k)^2\times SO(4k)^{\frac{n}{2}-1}\times Sp(2k)^{\frac{n}{2}-2}\times Sp(k)^2$ gauge theory, with bi-fundamental half-hypers (fig. \ref{fig6l} (b)), described by a $\mathbb{D}_{n-2}$ quiver. The $i$th factor of the gauge group has $v_i$ fundamentals, such that $\sum_i{v_i}l_i=n$, where $l_i=1,2$ is the Dynkin label for the $i$th node.

\begin{equation}
dim M^B_C=2 k (n-1)
\end{equation}
\begin{equation}
dim M^B_H=2 k (m-1)
\end{equation}

$\bullet$ \emph{n odd,m even (n{\tt \symbol{62}}3,m{\tt \symbol{62}}4)}

\begin{figure}[htbp]
\begin{center}
\includegraphics[height=3.0in]{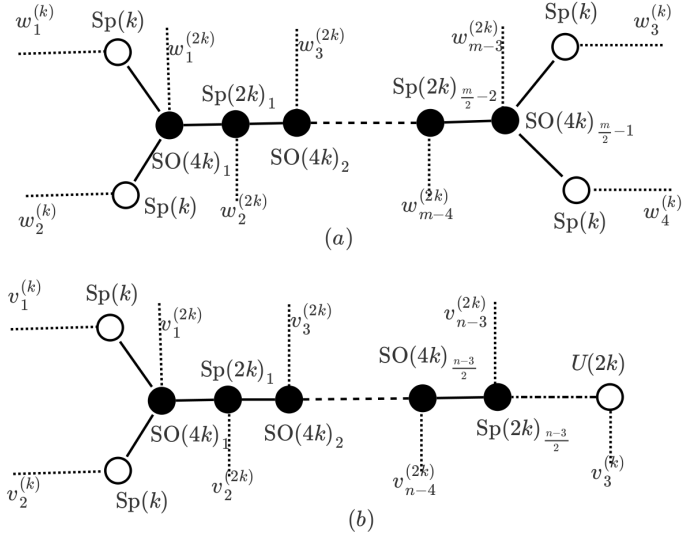}
\caption{Mirror Duals for $\Gamma_1=\mathbb{D}_{n-2}$,$\Gamma_2=\mathbb{D}_{m-2}$, with $n$ odd and $m$ even. The "dot-dash" line denotes the bifundamental hyper for  $Sp(2k)\times U(2k)$ in figure (b)}
\label{fig7l}
\end{center}
\end{figure}

A-model:$Sp(k)^2\times SO(4k)^{\frac{m}{2}-1}\times Sp(2k)^{\frac{m}{2}-2}\times Sp(k)^2$ gauge theory, with bi-fundamental half-hypers (fig. \ref{fig7l} (a)), described by a $\mathbb{D}_{m-2}$ quiver. The $i$th factor of the gauge group has $w_i$ fundamentals, such that $\sum_i{w_i}l_i=n$, where $l_i=1,2$ is the Dynkin label for the $i$th node.

\begin{equation}
dim M^A_C=2 k (m-1)
\end{equation}
\begin{equation}
dim M^A_H=2 k (n-1)
\end{equation}
B-model:$Sp(k)^2\times SO(4k)^{\frac{n-3}{2}}\times Sp(2k)^{\frac{n-3}{2}}\times U(2k)$ gauge theory, with half-hypers in the bi-fundamental of $Sp(k)\times SO(4k)$ and $Sp(2k)\times SO(4k)$ ( \ref{fig7l} (b)) and one hyper in the bi-fundamental of $Sp(2k)\times U(2k)$ (fig. \ref{fig7l} (b)). The $i$th factor of the gauge group has $v_i$ fundamentals, such that $\sum_i{v_i}l_i=n$, where $l_i=\frac{Rank(g_i)}{k}$ and $g_i$ is the factor in the gauge group. 
\begin{equation}
dim M^B_C=2 k (n-1)
\end{equation}
\begin{equation}
dim M^B_H=2 k (m-1)
\end{equation}

$\bullet$ \emph{n odd,m odd (n,m{\tt \symbol{62}}3)}

\begin{figure}[htbp]
\begin{center}
\includegraphics[height=2.0in]{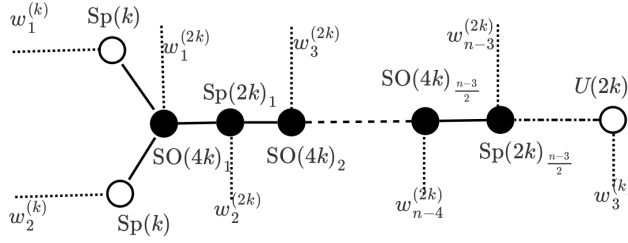}
\caption{A-model for $\Gamma_1=\mathbb{D}_{n-2}$,$\Gamma_2=\mathbb{D}_{m-2}$, with $n,m$ odd. B-model is identical with $m$ replaced by $n$ and the integers$\{w_i\}$s by $\{v_i\}$}
\label{fig8l}
\end{center}
\end{figure}

A-model:$Sp(k)^2\times SO(4k)^{\frac{m-3}{2}}\times Sp(2k)^{\frac{m-3}{2}}\times U(2k)$ gauge theory, with half-hypers in the bi-fundamental of $Sp(k)\times SO(4k)$ and $Sp(2k)\times SO(4k)$ (fig.\ref{fig8l} ) and one hyper in the bi-fundamental of $Sp(2k)\times U(2k)$ . The $i$th factor of the gauge group has $v_i$ fundamentals, such that $\sum_i{w_i}l_i=n$, where $l_i=\frac{Rank(g_i)}{k}$ and $g_i$ is the factor in the gauge group.

\begin{equation}
dim M^A_C=2 k (m-1)
\end{equation}
\begin{equation}
dim M^A_H=2 k (n-1)
\end{equation}
B-model:$Sp(k)^2\times SO(4k)^{\frac{n-3}{2}}\times Sp(2k)^{\frac{n-3}{2}}\times U(2k)$ gauge theory, with half-hypers in the bi-fundamental of $Sp(k)\times SO(4k)$ and $Sp(2k)\times SO(4k)$  and one hyper in the bi-fundamental of $Sp(2k)\times U(2k)$ . The $i$th factor of the gauge group has $v_i$ fundamentals, such that $\sum_i{v_i}l_i=n$, where $l_i=\frac{Rank(g_i)}{k}$ and $g_i$ is the factor in the gauge group. The quiver diagram for the B-model can simply be obtained by replacing $m$ and $n$ and the $w_i$s by $v_i$s in figure \ref{fig8l}.

\begin{equation}
dim M^B_C=2 k (n-1)
\end{equation}
\begin{equation}
dim M^B_H=2 k (m-1)
\end{equation}
The mirror duals resulting from (1),(2) and (3) have been discussed in the literature \cite{Intriligator:1996ex},\cite{deBoer:1996mp} in great detail. In the appendices, we derive the field content of the mirror dual theories in the cases (4)-(8), using standard orbifold/orientifold projection techniques \cite{Douglas:1996sw} directly from the Type IIA brane construction described above.

There are two points which need to be discussed for the theories involving D-type singularities:

\begin{enumerate}%
\item The map between the set of integers $\{w_i\}$ and $\{v_j\}$ is not clear in these cases.

\item There seems to be an apparent mismatch of the number of masses and FI parameters between the dual theories. For example, in the $\mathbb{Z}_n \times \mathbb{D}_{m-2}$ ($n$ even) theories, the B-model has $\frac{n}{2}-1$ FI parameters and it is not clear apriori whether this matches with the number of independent hypermultiplet masses in the A-model.

\end{enumerate}
In the next section, we shall address the above questions by resorting to Type IIB brane constructions. We shall restrict  our discussion to $\mathbb{Z}_n \times \mathbb{D}_{m-2}$- type mirror dual theories, for even $n$.\\

Type IIB description for some of the theories listed above have been discussed in \cite{Kapustin:1998fa}, \cite{Hanany:1999sj} and \cite{Feng:2000eq}.
Generic $Sp \times SO$ quivers ( recall that mirror duals arising out of $\mathbb{D}_{n-2} \times \mathbb{D}_{m-2}$ singularities in the M-theory are of this type) have been discussed in \cite{Gaiotto:2008ak} in some detail. The Type IIB brane construction in \cite{Gaiotto:2008ak} can be used to find the appropriate map between the sets of integers $\{w_i\}$ and $\{v_j\}$ in a way similar to the $\mathbb{Z}_n \times \mathbb{D}_{m-2}$ case, described below. However, we will leave a discussion of this particular category of mirror duals for a future work.

\subsection{$\mathbb{Z}_n \times \mathbb{D}_{m-2}$ theories : Type IIB description}

\begin{figure}[htbp]
\begin{center}
\includegraphics[height=3.0in]{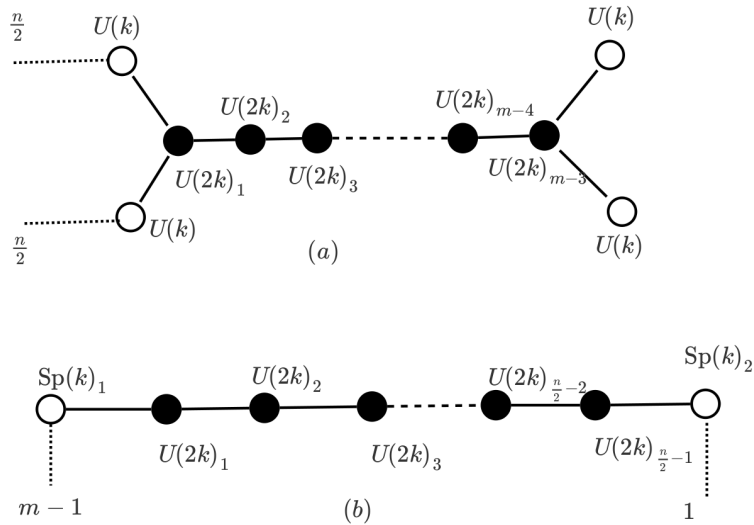}
\caption{A mirror pair of $\mathbb{Z}_n \times \mathbb{D}_{m-2}$ type, $n$ even}
\label{fig9l}
\end{center}
\end{figure}

\begin{figure}[htbp]
\begin{center}
\includegraphics[height=3.0in]{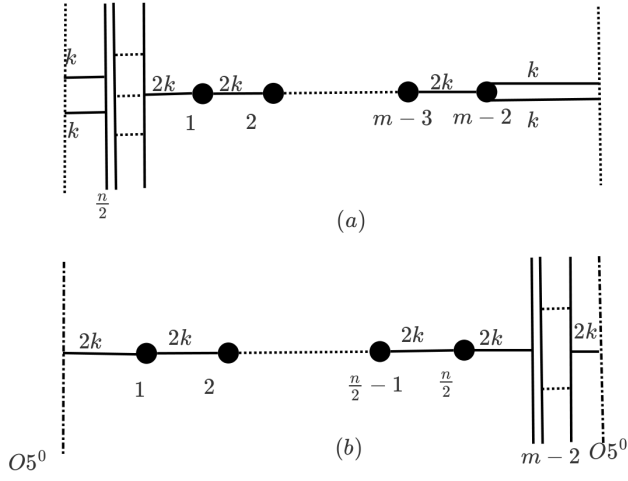}
\caption{Type IIB description of the mirror pairs in figure \ref{fig9l}}
\label{fig11l}
\end{center}
\end{figure}

Generically, Type IIB description of theories arising from D-type singularities in the M-theory requires additional ingredients - $O5$ planes and orbifold 5-planes \cite{Kapustin:1998fa}. S-duality maps the orbifolding operation $(-)^{F_L}\mathcal{I}_4$ into the orientifolding operation $\Omega \mathcal{I}_4$, where $\mathcal{I}_4$ denotes a $\mathbb{Z}_2$ action on 4 of the spatial directions (we will specify which ones below). For concreteness, consider the mirror pair in figure \ref{fig9l}. The A-model is essentially a $\mathbb{D}_{m-2}$ quiver gauge theory with $\frac{n}{2}$ hypers on two of the nodes with Dynkin label 1 (corresponding to gauge groups $U(k)_1$ and $U(k)_2$). The brane construction for this theory (shown in figure \ref{fig11l} (a)) involves the same background $\mathcal{M}:\mathbb{R}^{2,1}\times S^1 \times \mathbb{R}^3_{\vec{X}} \times \mathbb{R}^3_{\vec{Y}}$ with D3, D5 and NS5 branes extending in precisely the same directions as described in section 2. In  figure \ref{fig11l}, the black spheres and the solid vertical lines denote the NS5s and the D5s respectively, while the horizontal solid lines represent D3 branes. The vertical dotted lines represent orbifold fixed planes while vertical dot-dashed lines are used to represent $O5^{0}$ planes (to be defined below). The orbifold fixed planes are parallel to the NS5 branes while the $O5^{0}$ planes are parallel to the D5 branes.

We consider a $\mathbb{Z}_2$ orbifolding on $S^1 \times \mathbb{R}^3_{\vec{X}}$ - directions transverse to the D5 branes.The compact direction along the D3 branes is, therefore, the orbifolded circle $S^1/\mathbb{Z}_2$ and the fixed points (at $s=0,L$) correspond to the locations of the two orbifold fixed planes in the compact direction. We place $m-2$ NS5 branes in this interval with $2k$ D3 branes stretching between consecutive pair of NS5s. At each of the two boundaries, we also have $2k$ D3 branes connecting the orbifold fixed plane and the nearest NS5 brane. Finally, we put $\frac{n}{2}$ D5 branes in the interval between $s=0$ and the nearest NS5 branes. The particle-content of the resultant three-dimensional theory can now be obtained by standard orbifold-projection techniques. Away from the boundaries, open strings on the $2k$ D3 branes between any two consecutive NS5 branes give a $U(2k)$ vector multiplet while those connecting D3 branes stretched between adjoining pairs of NS5s give bifundamental hypermultiplets as before. But there is an important subtlety involving D3 branes which end on the orbifold fixed planes. These branes are charged under the RR twisted sector $U(1)$ gauge field living on the orbifold plane, as can be verified directly by a boundary state computation \cite{Sen:1998ii}. The computation also shows that ,for D3 branes ending on the fixed plane, there are two allowed configurations , differing in their charge under the twisted RR gauge field - we shall call these ''+'' and ``-'' states respectively.

Now, suppose we have $k_+$ D3 branes in the ''+'' state and $k_-$ D3 branes in the ``-'' state($k_+ + k_-=2k$). The action of orbifold $(-1)^{F_L} \mathcal{I}_4$ on the CP factors of the D3 branes is given by a diagonal matrix with $k_+$ entries equal to +1 and $k_-$ entries equal to -1. Obviously, this breaks the gauge group from $U(2k)$ to $U(k_+) \times U(k_-)$ and a similar action on the D5 CP factors (diagonal,$n/2$ entries +1 and $n/2$ entries -1) implies that each factor in $U(k_+) \times U(k_-)$ has 
$n/2$ fundamental hypers. The bi-fundamental hypers of $U(k_+) \times U(k_-)$ are projected out by the boundary conditions on the NS5 brane. Therefore, for $k_+=k_-=k$ at both boundaries, we obtain the required quiver gauge theory for the A-model.

The brane construction clearly shows that the A-model has a total of $\frac{n}{2}-1$ fundamental mass parameters - putting all the D3 branes at the $\mathbb{R}^3_{\vec{X}}$ position of the $i$th D5 brane leaves exactly $\frac{n}{2}-1$ free parameters. Since D3 branes on the left and right of each NS5 brane can be chosen to have same position in $\mathbb{R}^3_{\vec{X}}$, all the bifundamental masses can be set to zero. Also, the total number of FI parameters is $m-1$, which can be seen as follows. The FI parameters for the $U(k)_1$ and $U(k)_2$ gauge groups are obviously equal, since they are related to the $\mathbb{R}^3_{\vec{Y}}$ position of the same NS5. The same argument holds for the $U(k)_3$ and $U(k)_4$ gauge groups. The total number is therefore, $(m-3) +2=m-1$.

Now, mirror dual for this system of branes (shown in figure \ref{fig11l} (b)) is given by the action of S-duality (followed by a rotation $\mathbb{R}^3_{\vec{X}} \leftrightarrow \mathbb{R}^3_{\vec{Y}}$). The NS5s are mapped into D5s and vice-versa, while each orbifold fixed plane is mapped into a $O5^{0}$ plane - a single D5 brane coincident with a parallel $O5^{-}$ plane. Therefore, the B-model consists of $\frac{n}{2}$ NS5 branes placed between two $O5^{0}$ planes in the compact direction. In addition, there are $m-2$ D5 branes between one of the NS5 branes at the boundary and the closest $O5^{0}$ plane. This naturally leads to the quiver gauge theory of \ref{fig9l}(b) - a $Sp(k) \times U(2k)^{n/2-1}\times Sp(k)$ gauge theory (bifundamental hypers as indicated in the figure) with $m-1$ hypers in the fundamental of one of the $Sp(k)$s and 1 hyper in the fundamental of the other.

Evidently, the number of FI parameters for the B-model is $\frac{n}{2}-1$ while the number of fundamental mass parameters is $m-1$. Since the D3 branes at the two boundaries actually end on the boundary D5 branes and therefore have to be localized at the same $\mathbb{R}^3_{\vec{X}}$ points as the respective D5s, the two fundamental hypers (originating from the D3-D5 open strings for the boundary D5s) will be massless. The number of fundamental masses will then correspond to the positions of the remaining D5 branes, namely, $(m-2)$. However,  
all the bifundamental masses cannot  be set to zero by the same reasoning as before. Since the D3 branes at the two boundaries have different $\mathbb{R}^3_{\vec{X}}$ positions, one can only set $\frac{n}{2}$ of those to zero leaving behind one massive bifundamental hyper. This leads to  a total of $m-1$ mass parameters in the B-model.

Therefore, the exchange of FI parameters and masses in mirror duality can be established in this case as well, with the brane construction giving a clear map between the two sets of parameters. The derivation above also shows that there is no real mismatch in their numbers between the dual theories. The map between the sets of integers $\{w_i\}$ and $\{v_j\}$ encoding the distribution of fundamental hypers on the dual theories can also be read off from the rules of S-duality stated above. In the next section, we will compute the partition functions of the dual theories using localization techniques and explicitly derive the mirror map.

\section{Computation of Partition Functions for the Dual Theories}
\subsection{Localization on $S^3$ and Mirror Symmetry}

In the IR limit, one can compute the expectation values of a certain set of observables (described below)  in $\mathcal{N}=4$ theories in $d=3$ by evaluating the corresponding path integral on $S^3$ with localization techniques.

To localize the path integral of a theory (deformed by mass and FI parameters)on $S^3$, one first needs to identify the supercharge $Q$ corresponding to a Killing spinor $\epsilon$ on $S^3$. The prescription,then, is to add a $Q$-exact term to the action:

\begin{equation}
S_{loc}=t \int  d^3 x Q (\sum \{Q,\Psi_i\}^{\dagger} \Psi_i)
\end{equation}
where the sum extends over all fermions in the theory and $t$ is a real parameter. One can easily show that the expectation values of the partition function and other $Q-closed$ observables are invariant under such a transformation and hence can be calculated in the $t \rightarrow \infty$ limit. In this limit, the path integral localizes to field configurations which makes the $Q-exact$ functional vanish.

The zero locus of the functional requires all bosonic fields in the matter hypermultiplets to vanish. For the vector multiplets, the only non-vanishing fields are $\sigma, D$ living in the $\mathcal{N}=2$ vector multiplet (which is part of the $\mathcal{N}=4$ vector multiplet), such that $\sigma = -D = \sigma_0$, where $\sigma_0$ is a constant on $S^3$. Also, one can gauge-transform $\sigma_0$ to an element of the Cartan sub-algebra of the gauge group, by introducing a Vandermonde determinant into the integration measure of the path-integral. For the background vector-multiplets, one can show \cite{Kapustin:2010xq} that $\Phi_m=\hat{\Phi}_{FI}=0$, while $\hat{V}_{FI}$ and $V_m$ contribute respectively to the classical action and the 1-loop partition function as described below.

In the limit of large t, the saddle point approximation for the partition function is given as,

\begin{equation}
Z=\int  d\sigma_0 \exp{S_{cl}[\sigma_0]} Z_{1-loop}[\sigma_0]
\end{equation}
The classical and 1-loop contributions for the gauge and matter fields can be summarized as follows:

\textbf{Classical Action}

For the $\mathcal{N}=4$ theories with superconformal IR fixed points, $S_{cl}[\sigma_0]$ does not have the usual quadratic term in $\sigma_0$ - it only contributes a term linear in $\sigma_0$ coming from the BF coupling with the background vector multiplet $\hat{V}_{FI}$:

\begin{itemize}%
\item $S^{FI}_{cl}= 2\pi i \eta Tr(\sigma_0)$

\end{itemize}
for every $U(1)$ factor in the gauge group.

\textbf{1-loop Determinant}

\begin{itemize}%
\item Each $\mathcal{N}=4$ vector multiplet contributes

\begin{equation}
Z^v_{1-loop}=\prod_{\alpha} \frac{\sinh{\pi \alpha(\sigma_0)}}{(\pi \alpha(\sigma_0))^2}
\end{equation}
where the product extends over all the roots of the Lie algebra of G.

\item Each $\mathcal{N}=4$ hypermultiplet contributes

\begin{equation}
Z^h_{1-loop}=\prod_{\rho} \frac{1}{\cosh{\pi \rho(\sigma_0+m)}}
\end{equation}
where the product extends over all the weights of the representation R of the gauge group G and $m$ is a real mass parameter.
\end{itemize}

The Vandermonte factor in the measure exactly cancels with the denominator of the 1-loop contribution of the vector multiplet for each factor in the gauge group.

The integrand obtained from the above set of rules has to be divided by the order of the Weyl group to account for the residual gauge symmetry.

This completes the prescription for writing down the partition function on $S^3$ for any $\mathcal{N}=4$ gauge theory in three dimension with a given gauge group and a given matter content in some representation(s) of the gauge group (with the same set of restrictions as noted in \cite{Kapustin:2010xq} ).

In the next subsections, we will describe two examples in which mirror symmetry between two theories can be confirmed by computing the partition functions of individual theories and showing that they are identical, with masses in one theory becoming FI parameters in the other theory and vice-versa. 

\subsection{A Simple Example:$\Gamma_1=\mathbb{Z}_n,\Gamma_2=Trivial$}
The mirror duals for $\Gamma_1=\mathbb{Z}_n,\Gamma_2=Trivial$ are,\\
A-model: U(k) gauge theory with $n$ fundamental hypers and 1 adjoint hyper.\\
B-model: $U(k)^n$ gauge theory given by a $\mathbb{Z}_n$ quiver ($(n+1)$th node identified with the 1st) with one hyper in the fundamental of one of the $U(k)$s.\\

In the Type IIB (Hanany-Witten) picture, the A-model corresponds to the world volume theory of $k$ $D3$ branes beginning and ending on a single $NS5$ brane and $n$ $D5$ branes. The B-model, on the other hand, corresponds to sets of $k$ $D3$ branes stretched between $n$ $NS5$ branes and $1$ $D5$ brane.

From the set of rules summarized in the previous section, the partition functions of the A-model and the B-model on $S^3$ can be readily computed:

\begin{equation}
Z_A=\int  d^k\sigma \frac{\prod_{i < j}\sinh^2{\pi (\sigma^i-\sigma^j)}}{\prod_{i,j}\cosh{\pi(\sigma^i-\sigma^j+m_{adj})}\prod^{n}_{i;a=1}\cosh{\pi(\sigma^i+m^f_a)}} \exp{2 \pi i \eta \sum_i \sigma^i}
\end{equation}
\begin{equation}
Z_B=\int  \prod^n_{a=1} d^k\sigma_{a}\frac{1}{\prod_{i}\cosh{\pi (\sigma^i_1 +m)}}\frac{\prod_{i < j,a=1,..,n}\sinh^2{\pi(\sigma^i_a-\sigma^j_a)}}{\prod_{i,j,a=1,..,n}\cosh{\pi(\sigma^i_a-\sigma^j_{a+1}+m_a)}} \exp{2\pi \sum_{i,a} \eta_a \sigma^i_a}
\end{equation}
where $\sigma^i_{n+1}$ is identified with $\sigma^i_{1}$ .\\

Now, $Z_B$ (as also $Z_A$) can be recast into the following form:

\begin{equation}
\begin{split}
Z_B&= \int  \prod^{n+1}_{a=1} d^k\sigma_{a} d^k\tau_{a}(\sum_{\rho_1} (-1)^{\rho_1}\prod_{i}\frac{\exp{2 \pi i \tau^{i}_1(\sigma^i_1-\sigma^{\rho_1(i)}_2)}}{\cosh{\pi (\sigma^i_1 +m)}})\\
& \times (\prod^{n+1}_{b=2}  \sum_{\rho_b} (-1)^{\rho_b}\prod_{i}\frac{\exp{2 \pi i \tau^{i}_b(\sigma^i_b-\sigma^{\rho_b(i)}_{b+1} +m_b)}}{\cosh{\pi (\tau^i_b)}}
\exp{2\pi \sum_{i} \eta_b \sigma^i_b})
\end{split}
\end{equation}
where $\sigma^i_{n+2}$ is identified with $\sigma^i_1$. Following \cite{Kapustin:2010xq},  the expression $\sum_{\rho_1} (-1)^{\rho_1}\prod_{i}\frac{\exp{2 \pi i \tau^{i}_1(\sigma^i_1-\sigma^{\rho_1(i)}_2)}}{\cosh{\pi (\sigma^i_1 +m)}}$ can be interpreted as the contribution of the only $D5$ brane to the partition function. The expression $\sum_{\rho_b} (-1)^{\rho_b}\prod_{i}\frac{\exp{2 \pi i \tau^{i}_b(\sigma^i_b-\sigma^{\rho_b(i)}_{b+1} +m_b)}}{\cosh{\pi (\tau^i_b)}}$, on the other hand, can be interpreted as the contribution of the $b$-th $NS5$ brane to the partition function.

Now, consider the numerator of the $Z_B$ integrand:

$\prod^{n+1}_{a=1} \sum_{\rho_a} (-1)^{\rho_a} \prod_i \exp{2 \pi i \tau^{i}_a(\sigma^i_a-\sigma^{\rho_a(i)}_{a+1})}$ (with $\sigma_{n+2}=\sigma_1$) \\
$=\prod^{n+1}_{a=1} \sum_{\rho_a} (-1)^{\rho_a} \prod_i \exp{2 \pi i \sigma^{i}_a(\tau^i_a-\tau^{\rho^{-1}_{a-1}(i)}_{a-1})}$(with $\tau^i_0=\tau^i_{n+1}$)\\
$\to \prod^{n+1}_{a=1} \sum_{\rho_a} (-1)^{\rho_a} \prod_i \exp{2 \pi i \sigma^{i}_a(\tau^i_a-\tau^{\rho_{a}(i)}_{a+1})}$.

In the last step, we relabel $\tau_a \rightarrow -\tau_{a+1}$, with $\tau_{n+2}=\tau_1$. The above equalities prove that the numerator of the $Z_B$ integrand has a symmetry under $\sigma_a \Longleftrightarrow -\tau_a$.

Given the said symmetry, the partition function $Z_B$ can be written as,

\begin{equation}
\begin{split}
Z_B&=\int  (\prod^{n+1}_{a=1} \sum_{\rho_a} (-1)^{\rho_a} \prod_i \exp{2 \pi i \sigma^{i}_a(\tau^i_a-\tau^{\rho_{a}(i)}_{a+1})})\frac{1}{\prod_{i}\cosh{\pi (\sigma^i_1 +m)}}\\
& \times(\prod_{a \neq 2}\frac{1}{\prod_{i}\cosh{\pi \tau^i_a}}) (\prod_{a \neq 1} e^{2 \pi i \sum_{i}\eta_a \sigma^i_a}  e^{2 \pi i \sum_{i}m_a \tau^i_a})
\end{split}
\end{equation}
Note that in the denominator we have implemented the redefinition $\tau_a \rightarrow -\tau_{a+1}$, which is necessary for making the above-mentioned symmetry of the numerator manifest.

The action of S-duality on the integration variables in the partition function is then given by $\sigma_a \Longleftrightarrow -\tau_a$. The S-dual partition function can be written as,

\begin{equation}
\begin{split}
\tilde{Z_B}&=\int  (\prod^{n+1}_{a=1} \sum_{\rho_a} (-1)^{\rho_a} \prod_i \exp{2 \pi i \tau^{i}_a(\sigma^i_a-\sigma^{\rho_{a}(i)}_{a+1})})\frac{1}{\prod_{i}\cosh{\pi (-\tau^i_1 +m)}}\\ 
& \times(\prod_{a \neq 2}\frac{1}{\prod_{i}\cosh{\pi \sigma^i_a}})(\prod_{a \neq 1} e^{-2 \pi i \sum_{i}\eta_a \tau^i_a}  e^{-2 \pi i \sum_{i}m_a \sigma^i_a})
\end{split}
\end{equation}
On integrating out the $\tau_{n+1},\sigma_{n+1}$;$\tau_{n},\sigma_{n}$;\ldots{}\ldots{}$\tau_{2},\sigma_{2}$, we have,

\begin{equation}
\begin{split}
\tilde{Z_B}&=\int  \sum_{{\rho_a}} (-1)^{\rho_1} (-1)^{\rho_2}.... (-1)^{\rho_{n+1}} \prod_i \exp{2 \pi i \tau^i_1(\sigma^i_1-\sigma^{\rho_{n+1}\rho_{n}....{\rho_2}{\rho_1}(i)}_1-\eta_2 -....-\eta_{n+1})}\\
&\times \frac{1}{\prod_{i}\cosh{\pi (-\tau^i_1 +m)}}\frac{1}{\prod_{i}\cosh{\pi (\sigma^i_1+\eta_3 +....+\eta_{n+1})}}\frac{1}{\prod_{i}\cosh{\pi (\sigma^i_1+\eta_4 +....+\eta_{n+1})}}......\\
&\times \frac{1}{\prod_{i}\cosh{\pi (\sigma^i_1+\eta_{n+1})}}\frac{1}{\prod_{i}\cosh{\pi (\sigma^i_1)}}\\
&\times e^{(-2 \pi i (m_2+m_3+...+m_{n+1})\sum_i \sigma^i_1)}
\end{split}
\end{equation}
The permutations $\rho_1,\rho_2,.....,\rho_{n+1}$ can all be combined into a single permutation since the integrand only depends on the product of their action on the index $i$. Then, integrating out $\tau_1^i$ and using the Cauchy determinant formula, we have,

\begin{equation}
\begin{split}
\tilde{Z_B}&=\int  d^k\sigma_{1} \frac{\prod_{i < j} \sinh^2{\pi(\sigma^i_{1}-\sigma^j_{1})}}{\prod_{i,j}\cosh{\pi(\sigma^i_{1}-\sigma^j_{1}-\eta_2-...-\eta_{n+1})}}\\
& \times\frac{1}{\prod_i \cosh{\pi(\sigma^i_{1})} \prod_i \cosh{\pi(\sigma^i_{1}+\eta_{n+1})}....
\prod_i \cosh{\pi(\sigma^i_{1}+\eta_{n+1}+....+\eta_{3})}}\\
& \times e^{(-2\pi i(m_2+m_3+....+m_{n+1})\sum_i \sigma^i_{1})}
\end{split}
\end{equation}
We note that the above expression is identical to $Z_A$, provided the parameters in the two expressions are related in a particular way. This set of linear equations gives the appropriate mirror map for the dual theories in question.\\

\textbf{Mirror Map} Comparing the expressions for $Z_A$ and $Z_B$, one can read off the mirror map, as follows:

\begin{equation}
m_{ad}=-(\eta_2 + \eta_3...+ \eta_{n+1})
\end{equation}
\begin{equation}
m^f_1=0,m^f_2=\eta_{n+1},......,m^f_n=\eta_3+\eta_4+....+\eta_{n}+\eta_{n+1},
\end{equation}
\begin{equation}
\eta=-(m_2+m_3+....+m_{n+1})
\end{equation}
By a suitable relabeling of the FI parameters of the B-model, one can rewrite the mirror map in the more conventional form,

\begin{equation}
\eta_i=m^f_{i+1}-m^f_{i} (i < n),\eta_n=m^f_1 -m^f_{n}- m_{ad}
\end{equation}
\subsection{Partition Functions of Mirror Duals :$\mathbb{Z}_n \times \mathbb{D}_{m-2}$ Case}

In this section, we compute the partition functions for mirror dual theories arising from $\mathbb{Z}_n \times \mathbb{D}_{m-2}$ singularities in the M-theory picture and explicitly obtain the mirror map for the case where $n$ is even. For concreteness, we consider the particular example of mirror duals in this category studied in the previous section (figure \ref{fig9l}), with the following field content:\\

A-model: A $U(k)^2 \times U(2k)^{m-3} \times U(k)^2$ gauge theory with bi-fundamental hypers and two sets of $\frac{n}{2}$ hypers in the fundamental of $U(k)_1$ and $U(k)_2$ respectively (figure \ref{fig9l} (a)). \\

B-model: A $Sp(k) \times U(2k)^{\frac{n}{2}-1} \times Sp(k)$ with bi-fundamental hypers and $(m-1)$ hypers in the fundamental of $Sp(k)_1$ and one hyper in the fundamental of $Sp(k)_2$ (figure \ref{fig9l} (b)).\\

The partition functions of the above theories on $S^3$ in the IR limit can be readily obtained using the rules summarized in the previous section.

\begin{equation}
\begin{split}
Z_A&=\int  d\sigma_{\alpha}d\tilde{\sigma}_{\beta} \frac{1}{\prod^{n/2}_{i,a=1}\cosh{\pi(\sigma_1^i+m^f_a)} \prod^{n/2}_{i;a=1}\cosh{\pi(\sigma_2^i+m^f_a)}}\frac{\prod_{i < j,\alpha=1,..,4} \sinh^2{\pi(\sigma^i_{\alpha}-\sigma^j_{\alpha})}}{\prod_{i,p;(\alpha,\beta)}\cosh{\pi(\sigma^i_{\alpha}-\tilde{\sigma}^p_{\beta}+m_{\alpha\beta})}}\\
&\times \frac{\prod_{p < l,\beta=1,..,m-3} \sinh^2{\pi(\tilde{\sigma}^p_{\beta}-\tilde{\sigma}^l_{\beta})}}{\prod_{p,l;\beta=1,2,...,m-4}\cosh{\pi(\tilde{\sigma}^p_{\beta}-\tilde{\sigma}^l_{\beta+1}+m_{\beta})}} \prod_{\alpha}e^{2\pi i \eta_{\alpha}\sum_i \sigma^i_{\alpha}} \prod_{\beta}e^{2\pi i \tilde{\eta}_{\beta}\sum_p\tilde{\sigma}^p_{\beta}}  
\end{split}
\end{equation}
where the adjoint scalars for $U(k)_{\alpha}$ is denoted as $\sigma^i_{\alpha}$ and those for $U(2k)_{\beta}$ as $\tilde{\sigma}^p_{\beta}$. $(\alpha,\beta)$ labels the hypermultiplet  in the bi-fundamental of $U(k)_{\alpha}\times U(2k)_{\beta}$ - obviously, $\beta=1$ for $\alpha=1,2$ and 
$\beta=m-3$ for $\alpha=3,4$.
\begin{equation}
\begin{split}
Z_B&=\int  d\sigma_{\alpha}d\sigma^{'}_{\beta} \Pi^{m-1}_{a=1}\frac{1}{\Pi_{i}\cosh{\pi(\sigma_1^i+M^f_a)}\cosh{\pi(\sigma_1^i-M^f_a)}}
\frac{1}{\Pi_{i}\cosh^2{\pi \sigma_{\frac{n}{2}}^i}} F(\sigma_1,\sigma^{'}_1;m_{bif}) F(\sigma_{\frac{n}{2}},\sigma^{'}_{\frac{n}{2}-1};m^{'}_{bif})\\
&\times\frac{\Pi_{p < l,\beta=1,..,\frac{n}{2}-2}\sinh{\pi(\sigma^{'p}_{\beta}-\sigma^{'l}_{\beta})}\sinh{\pi(\sigma^{'p}_{\beta+1}-\sigma^{'l}_{\beta+1})}}{\Pi_{i,j,\beta=1,..,\frac{n}{2}-2}\cosh{\pi(\sigma^{'p}_{\beta}-\sigma^{'l}_{\beta+1}+M_{\beta})}} \Pi^{\frac{n}{2}-1}_{\beta=1} e^{2\pi i \eta^{'}_{\beta}\Sigma_p \sigma^{'p}_{\beta}}
\end{split}
\end{equation}
where the adjoint scalars for the two $Sp(k)$ factors are denoted as $\sigma^i_1$ and $\sigma^i_{n/2}$ respectively, while the scalars in the adjoint of $U(2k)_{\beta}$ are
denoted as $\sigma^{'p}_{\beta}$. Also,
\begin{equation}
F(x,y;m_{xy})=\frac{\prod_{i < j} \sinh^2{\pi(x^i-x^j)}\sinh^2{\pi(x^i+x^j)} \prod_i \sinh^2{\pi 2 x^i } \prod_{p < l} \sinh{\pi(y^p-y^l)}}{\prod_{i,p}\cosh{\pi(x^i-y^p+m_{xy})}\cosh{\pi(x^i+y^p-m_{xy})}}
\end{equation}

In the expression for $Z_A$, we have set the masses of the fundamental hypers to be pairwise equal and one can easily check that this is a necessary condition for FI-like contributions corresponding to the $Sp(k)_1$ group to drop out from the dual partition function. Moreover, note that in the A-model the total number of independent fundamental mass parameters is $\frac{n}{2}-1$. This can be directly seen from $Z_A$ by redefining $\sigma^i_1 \rightarrow \sigma^i_1 -m^f_1$ and $\sigma^i_2 \rightarrow \sigma^i_2 -m^f_1$, which shifts the fundamental masses (and two of the bi-fundamental masses) by $-m^f_1$ and leaves a total of $\frac{n}{2}-1$ independent fundamental mass parameters.  This is consistent with the Hanany-Witten description of the A-model (discussed in the last section) which states that there should be $n/2$ $D5$ branes and hence $n/2 -1$ independent fundamental hyper masses .
Similarly one can show that the masses of the $U(k)_1 \times U(2k)_1$ and $U(k)_2 \times U(2k)_1$ bifundamental hypers have to be equal. The same is true for the $U(k)_3 \times U(2k)_{m-3}$ and the $U(k)_4 \times U(2k)_{m-3}$ bifundamental masses.
Our goal is to demonstrate that the partition functions listed above are related by a simple redefinition of integration variables, implying that the quiver gauge theories (A and B model) flow to the same superconformal theory in the IR.

We start with the A-model partition function.

Define $\tilde{\sigma}^p_0=(\sigma^i_1,\sigma^i_2)$,$\tilde{\sigma}^p_{m-2}=(\sigma^i_3,\sigma^i_4)$

Now, with this redefinition, $\frac{\prod_{i < j,\alpha=1,..,4} \sinh^2{\pi(\sigma^i_{\alpha}-\sigma^j_{\alpha})}}{\prod_{i,p;(\alpha,\beta)}\cosh{\pi(\sigma^i_{\alpha}-\tilde{\sigma}^p_{\beta}+m_{\alpha\beta})}}\frac{\prod_{p < l,\beta=1,..,m-3} \sinh^2{\pi(\tilde{\sigma}^p_{\beta}-\tilde{\sigma}^l_{\beta})}}{\prod_{p,l;\beta=1,...,m-4}\cosh{\pi(\tilde{\sigma}^p_{\beta}-\tilde{\sigma}^l_{\beta+1}+m_{\beta})}}$

$=X(\sigma_1,\sigma_2)(\frac{\prod_{p < l,\beta=0,1,..,m-3} \sinh{\pi(\tilde{\sigma}^p_{\beta}-\tilde{\sigma}^l_{\beta})}\sinh{\pi(\tilde{\sigma}^p_{\beta+1}-\tilde{\sigma}^l_{\beta+1})}}{\prod_{p,l;\beta=0,1,...,m-3}\cosh{\pi(\tilde{\sigma}^p_{\beta}-\tilde{\sigma}^l_{\beta+1}+m_{\beta})}}) X(\sigma_3,\sigma_4)$,

where $X(x,y)=\frac{\prod_{i < j} \sinh{\pi(x^i-x^j)}\sinh{\pi(y^i-y^j)}}{\prod_{i, j}\sinh{\pi(x^i-y^j)}}=\sum_{\rho} (-1)^{\rho}\frac{1}{\prod_i \sinh{\pi(x^i-y^{\rho(i)})}}$.

Thus, \\
$X(\sigma_1,\sigma_2)=\sum_{\rho} (-1)^{\rho}\frac{1}{\prod_i \sinh{\pi(\sigma^i_1-\sigma^{\rho(i)}_2)}}=\int  d\tau \sum_{\rho} (-1)^{\rho}\frac{\prod_i \sinh{\pi \tau^i}}{\prod_i \cosh{\pi \tau^i}} \exp{2 \pi i \tau^i (\sigma^i_1-\sigma^{\rho(i)}_2)}$.\\
$X(\sigma_3,\sigma_4)=\sum_{\rho^{'}}  (-1)^{\rho^{'}}\frac{1}{\prod_i \sinh{\pi(\sigma^i_3-\sigma^{\rho^{'}(i)}_4)}}=\int  d\tau^{'} \sum_{\rho^{'}} (-1)^{\rho^{'}}\frac{\prod_i \sinh{\pi \tau^{'i}}}{\prod_i \cosh{\pi \tau^{'i}}} \exp{2 \pi i \tau^{'i} (\sigma^i_3-\sigma^{\rho(i)}_4)}$.

The $NS5$ contribution in the partition function $Z_A$ can be rewritten as,

\begin{equation}
\begin{split}
&\frac{\prod_{p < l,\beta=0,1,..,m-3} \sinh{\pi(\tilde{\sigma}^p_{\beta}-\tilde{\sigma}^l_{\beta})}\sinh{\pi(\tilde{\sigma}^p_{\beta+1}-\tilde{\sigma}^l_{\beta+1})}}{\prod_{p,l;\beta=0,1,...,m-3}\cosh{\pi(\tilde{\sigma}^p_{\beta}-\tilde{\sigma}^l_{\beta+1}+m_{\beta})}}\\
&=\int  \prod_{\beta=0,1,..,m-3}(\sum_{\rho_{\beta}} (-1)^{\rho_{\beta}}\prod_p \frac{\exp{2 \pi i \tilde{\tau}^p_{\beta}(\tilde{\sigma}^p_{\beta}-\tilde{\sigma}^{\rho_{\beta}(p)}_{\beta+1}+m_{\beta})}}{\cosh{\pi \tilde{\tau}^p_{\beta}}}) d\tilde{\tau}^p_{\beta}
\end{split}
\end{equation}
Combining the $NS5$ contribution in the partition function with $X(\sigma_3,\sigma_4)$, we have,

\begin{equation}
\begin{split}
T_{NS5}&=\int  \prod_{p,\beta}d\tilde{\tau}^p_{\beta} \prod_i d\tau^{'i} (\prod_{\beta=0,1,..,m-3} \sum_{\rho_{\beta}} (-1)^{\rho_{\beta}}\prod_p \frac{\exp{2 \pi i \tilde{\tau}^p_{\beta}(\tilde{\sigma}^p_{\beta}-\tilde{\sigma}^{\rho_{\beta}(p)}_{\beta+1}+m_{\beta})}}{\cosh{\pi \tilde{\tau}^p_{\beta}}})\\
& \times(\sum_{\rho^{'}} (-1)^{\rho^{'}}\frac{\prod_i \sinh{\pi \tau^{'i}}}{\prod_i \cosh{\pi \tau^{'i}}} \exp{2 \pi i \tau^{'i} (\sigma^{'i}_{m-2}-\sigma^{k+\rho(i)}_{m-2})}) \prod_{\alpha}e^{2\pi i \eta_{\alpha}\sum_i \sigma^i_{\alpha}} \prod_{\beta}e^{2\pi i \tilde{\eta}_{\beta}\sum_p\tilde{\sigma}^p_{\beta}} 
\end{split}
\end{equation}
Similarly,combining the ''$D5$'' contribution (i.e. the fundamental hypers) in the partition function with $X(\sigma_1,\sigma_2)$, we have,

\begin{equation}
\begin{split}
T_{D5}&=\frac{1}{\prod_{i,a}\cosh{\pi(\sigma_1^i+m^f_a)} \prod_{i,a}\cosh{\pi(\sigma_2^i+m^f_a)}} X(\sigma_1,\sigma_2)\\
&=\int  \prod^{n/2+1}_{p;a=1}d\tilde{\tau}^{'p}_{a} \prod_i d\tau^{i} (\prod^{\frac{n}{2}}_{a=1} \sum_{\rho_a}(-1)^{\rho_a}\prod_p \frac{\exp{2 \pi i \tau^{'p}_a (\sigma^{'p}_a-\sigma^{'\rho_a(p)}_{a-1})}}{\cosh{\pi(\sigma_a^{'p}+m^f_a)}})\\
&\times (\sum_{\rho_{\frac{n}{2}+1}}(-1)^{\rho_{\frac{n}{2}+1}}\exp{2 \pi i \tau^{'p}_{\frac{n}{2}+1} (\sigma^{'p}_{\frac{n}{2}+1}-\sigma'^{\rho_{\frac{n}{2}+1}(p)}_{\frac{n}{2}})}) \\
&\times(\sum_{\rho} (-1)^{\rho}\frac{\prod_i \sinh{\pi \tau^i}}{\prod_i \cosh{\pi \tau^i}} \exp{2 \pi i \tau^i (\sigma^{'i}_{\frac{n}{2}+1}-\sigma^{k+\rho(i)}_{\frac{n}{2}+1})})
\end{split}
\end{equation}
with the identification $\sigma^{'p}_{0}=\tilde{\sigma}^p_0=(\sigma^i_1,\sigma^i_2)$. Note that we have added an extra pair of auxiliary variables $(\sigma_{\frac{n}{2}+1},\tau_{\frac{n}{2}+1})$ in the integral.

Since the $NS5$ contribution is antisymmetric under permutation of the indices of $\tilde{\sigma}^p_0$, the $D5$ contribution can be antisymmetrized for $\tilde{\sigma}^p_0 (i.e. \sigma^{'p}_{0})$ and by the same reasoning, for $\sigma^{'p}_{1},\sigma^{'p}_{2},.....,\sigma^{'p}_{\frac{n}{2}}$, as we have done above.

Putting the two contributions together, we have,

\begin{equation}
\begin{split}
Z_A&=\int (\sum_{\rho} (-1)^{\rho}\frac{\prod_i \sinh{\pi \tau^i}}{\prod_i \cosh{\pi \tau^i}} \exp{2 \pi i \tau^i (\sigma^{'i}_{\frac{n}{2}+1}-\sigma^{k+\rho(i)}_{\frac{n}{2}+1})})(\prod^{\frac{n}{2}+1}_{a=1} \sum_{\rho_a}(-1)^{\rho_a}\prod_p \frac{\exp{2 \pi i \tau^{'p}_a (\sigma^{'p}_a-\sigma^{'\rho_a(p)}_{a-1})}}{I_a(\sigma^{'},\tau^{'})})\\
&\times(\prod_{\beta=0,1,..,m-3} \sum_{\rho_{\beta}} (-1)^{\rho_{\beta}}\prod_p \frac{\exp{2 \pi i \tilde{\tau}^p_{\beta}(\tilde{\sigma}^p_{\beta}-\tilde{\sigma}^{\rho_{\beta}(p)}_{\beta+1}+m_{\beta})}}{\cosh{\pi \tilde{\tau}^p_{\beta}}})\\ 
&\times(\sum_{\rho^{'}} (-1)^{\rho^{'}}\frac{\prod_i \sinh{\pi \tau^{'i}}}{\prod_i \cosh{\pi \tau^{'i}}} \exp{2 \pi i \tau^{'i} (\sigma^{'i}_{m-2}-\sigma^{k+\rho(i)}_{m-2})}) \\
&\times \prod_{\alpha}e^{2\pi i \eta_{\alpha}\sum_i \sigma^i_{\alpha}} \prod_{\beta}e^{2\pi i \tilde{\eta}_{\beta}\sum_p\tilde{\sigma}^p_{\beta}} 
\end{split}
\end{equation}
where $I_a(\sigma^{'},\tau^{'})=1,\cosh{\pi (\sigma^{'p}_a+m^f_a)}$ for $a=\frac{n}{2}+1, a \leq \frac{n}{2}$ respectively.

We now demonstrate that the action of S-duality on the integration variables is given by the simple redefinition: $\sigma^{'p} \leftrightarrow -\tau^{'p},\tilde{\sigma}^p \leftrightarrow -\tilde{\tau}^p$; $\tau,\tilde{\tau}$ remaining invariant. To this end, we will consider the action of S-duality on the individual terms $T_{D5}$ and $T_{NS5}$.

In $T_{D5}$, the numerator transforms as follows under the transformation $\sigma^{'p} \leftrightarrow -\tau^{'p}$,

\begin{equation}
\begin{split}
 \prod^{\frac{n}{2}+1}_{a=1} \sum_{\rho_a}(-1)^{\rho_a}\prod_p \exp{2 \pi i \tau^{'p}_a (\sigma^{'p}_a-\sigma^{'\rho_a(p)}_{a-1})}& \rightarrow \prod^{\frac{n}{2}+1}_{a=1} \sum_{\rho_a}(-1)^{\rho_a}\prod_p \exp{2 \pi i \sigma^{'p}_a (\tau^{'p}_a-\tau^{'\rho_a(p)}_{a-1})}\\
&= \prod^{\frac{n}{2}+1}_{a=1}\sum_{\rho_a}(-1)^{\rho_a}\prod_p \exp{2 \pi i \tau^{'p}_a (\sigma^{'p}_a-\sigma^{'\rho^{-1}_{a+1}(p)}_{a+1})} \\
&\times\exp{2 \pi i \sigma^{'p}_{\frac{n}{2}+1} \tau^{'p}_{\frac{n}{2}+1}} \exp{-2 \pi i \sigma^{'p}_1 \tau^{'\rho_1(p)}_0}.
\end{split}
\end{equation}
Hence,

\begin{equation}
\begin{split}
\tilde{T}_{D5}&=\int  \prod^{\frac{n}{2}}_{a=1} \sum_{\rho_a}(-1)^{\rho_{a+1}}\prod_p \frac{\exp{2 \pi i \tau^{'p}_a (\sigma^{'p}_a-\sigma^{'\rho^{-1}_{a+1}(p)}_{a+1})}}{\cosh{\pi(-\tau^{'p}_a+m^f_a)}}\\ &\times(\sum_{\rho} (-1)^{\rho}\frac{\prod_i \sinh{\pi \tau^i}}{\prod_i \cosh{\pi \tau^i}} \exp{2 \pi i \tau^{'i}_{\frac{n}{2}+1} (\sigma^{'i}_{\frac{n}{2}+1}-\tau^i)} \exp{2 \pi i \tau^{'k+i}_{\frac{n}{2}+1} (\sigma^{'i}_{\frac{n}{2}+1}+\tau^{\rho^{-1}(i)})})\\
&\times(\sum_{\rho_1}(-1)^{\rho_1} \prod_p \exp{-2 \pi i \sigma^{'p}_1 \tau^{'\rho_1(p)}_0})
\end{split}
\end{equation}
On integrating out $\tau^{'p}_a$ (for  $a=1,2,...,\frac{n}{2})$ and $\tau^{'}$ (none of which appear in the $NS5$ term), we have,

\begin{equation}
\begin{split}
\tilde{T}_{D5}&=\int  \prod^{\frac{n}{2}}_{a=1} (\sum_{\rho_a}(-1)^{\rho_{a+1}}\prod_p \frac{\exp{2 \pi i m^{f}_a (\sigma^{'p}_a-\sigma^{'\rho^{-1}_{a+1}(p)}_{a+1})}}{\cosh{\pi(\sigma^{'p}_a-\sigma^{'\rho^{-1}_{a+1}(p)}_{a+1})}}) (\sum_{\rho} (-1)^{\rho}\frac{\prod_i \sinh{\pi \tau^i}}{\prod_i \cosh{\pi \tau^i}} \delta(\sigma^{'k+i}_{\frac{n}{2}+1}+\sigma^{'\rho^{-1}(i)}_{\frac{n}{2}+1}))\\
& \times (\sum_{\rho_1}(-1)^{\rho_1} \prod_p \exp{-2 \pi i \sigma^{'p}_1 \tau^{'\rho_1(p)}_0})
\end{split}
\end{equation}
Finally, using the Cauchy determinant formula, we obtain,

\begin{equation}
\begin{split}
\tilde{T}_{D5}&=\int  \prod^{\frac{n}{2}}_{a=1} \frac{\prod_{p < l}\sinh{\pi(\sigma^{'p}_{a}-\sigma^{'l}_{a})}\sinh{\pi(\sigma^{'p}_{a+1}-\sigma^{'l}_{a+1})}}{\prod_{p,l}\cosh{\pi(\sigma^{'p}_{a}-\sigma^{'l}_{a+1})}}
(\sum_{\rho} (-1)^{\rho}\frac{\prod_i \sinh{\pi 2 \sigma^{'i}_{\frac{n}{2}+1}}}{\prod_i \cosh^2{\pi\sigma^{'i}_{\frac{n}{2}+1}  }} \delta(\sigma^{'k+i}_{\frac{n}{2}+1}+\sigma^{'\rho^{-1}(i)}_{\frac{n}{2}+1}))\\
& \times (\sum_{\rho_1}(-1)^{\rho_1} \prod_p \exp{-2 \pi i \sigma^{'p}_1 \tau^{'\rho_1(p)}_0}) (\prod^{\frac{n}{2}}_{p;a=1} \exp{2 \pi i m^{f}_a (\sigma^{'p}_a-\sigma^{'\rho^{-1}_{a+1}(p)}_{a+1})})
\end{split}
\end{equation}
with $\tau^{'p}_0=\tilde{\tau}^p_0$.

Now, similarly, the numerator of the $NS5$ term transforms as,

\begin{equation}
\begin{split}
&\prod_{\beta=0,1,..,m-3}(\sum_{\rho_{\beta}} (-1)^{\rho_{\beta}}\prod_p \exp{2 \pi i \tilde{\tau}^p_{\beta}(\tilde{\sigma}^p_{\beta}-\tilde{\sigma}^{\rho_{\beta}(p)}_{\beta+1}+m_{\beta})}) \\
&\rightarrow 
\prod_{\beta=0,1,..,m-3}(\sum_{\rho_{\beta}} (-1)^{\rho_{\beta}}\prod_p \exp{2 \pi i \tilde{\sigma}^p_{\beta}(\tilde{\tau}^p_{\beta}-\tilde{\tau}^{\rho_{\beta}(p)}_{\beta+1}+m_{\beta})})\\
&=\prod_{\beta=1,..,m-3}(\sum_{\rho_{\beta-1}} (-1)^{\rho_{\beta-1}}\prod_p \exp{2 \pi i \tilde{\tau}^p_{\beta}(\tilde{\sigma}^p_{\beta}-\tilde{\sigma}^{\rho^{-1}_{\beta-1}(p)}_{\beta-1})})\prod_p \exp{2 \pi i \tilde{\sigma}^{p}_{0} \tilde{\tau}^{p}_0} \exp{-2 \pi i \tilde{\sigma}^{p}_{m-3} \tilde{\tau}^{p}_{m-2}} \\
&\times \prod^{m-3}_{\beta=0} \exp{-2 \pi i \tilde{\sigma}^{p}_{\beta} m_{\beta}}
\end{split}
\end{equation}
Therefore, one can write the $NS5$ contribution as,

\begin{equation}
\begin{split}
\tilde{T}_{NS5}&=\int  \prod_{\beta=1,..,m-3}(\sum_{\rho_{\beta-1}} (-1)^{\rho_{\beta-1}}\prod_p \frac{\exp{2 \pi i \tilde{\tau}^p_{\beta}(\tilde{\sigma}^p_{\beta}-\tilde{\sigma}^{\rho^{-1}_{\beta-1}(p)}_{\beta-1}-\tilde{\eta}_{\beta})}}{\cosh{\pi \tilde{\sigma}^p_{\beta}}})(\prod_p \frac{\exp{2 \pi i (\tilde{\sigma}^{p}_{0}-\eta^p_0 )\tilde{\tau}^{p}_0}}{\cosh{\pi \tilde{\sigma}^p_{0}}}) \\
&\times(\sum_{\rho_{m-3}} (-1)^{\rho_{m-3}} \prod_p\exp{-2 \pi i (\tilde{\sigma}^{\rho^{-1}_{m-3}(p)}_{m-3}+\eta^p_{m-2} )\tilde{\tau}^{p}_{m-2}}) \\
&\times (\sum_{\rho^{'}} (-1)^{\rho^{'}}\prod_i \frac{\sinh{\pi \tau^{'i}}}{\prod_i \cosh{\pi \tau^{'i}}} \exp{2 \pi i \tau^{'i} (\tilde{\tau}^{i}_{m-2}-\tilde{\tau}^{k+\rho^{'}(i)}_{m-2})}) \\
&\times(\prod^{m-3}_{p;\beta=0} \exp{-2 \pi i \tilde{\sigma}^{p}_{\beta} m_{\beta}})
\end{split}
\end{equation}
Integrating out $\tilde{\tau}^p_{1},\tilde{\tau}^p_{2},.......,\tilde{\tau}^p_{m-3}$, we have,

\begin{equation}
\begin{split}
\tilde{T}_{NS5}&=\int  \prod_{\beta=1,..,m-3}\prod_p \frac{1}{\cosh{\pi (\tilde{\sigma}^p_{0}+\sum^{\beta}_{m-3} \eta_{\beta})}}\\
&\times(\sum_{{\rho_{\beta}}}\prod_{\beta=0,1,...,m-3}(-1)^{\rho_{\beta}}\prod_p \exp{-2 \pi i (\tilde{\sigma}^{\rho^{-1}_{0}....\rho^{-1}_{m-3}(p)}_{0}+\eta+\eta^p_{m-2}) \tilde{\tau}^{p}_{m-2}})\\
&\times(\prod_p \frac{\exp{2 \pi i (\tilde{\sigma}^{p}_{0}-\eta^p_0)\tilde{\tau}^{p}_0}}{\cosh{\pi \tilde{\sigma}^p_{0}}})(\sum_{\rho^{'}} (-1)^{\rho^{'}}\prod_i \frac{\sinh{\pi \tau^{'i}}}{\prod_i \cosh{\pi \tau^{'i}}} \exp{2 \pi i \tau^{'i} (\tilde{\tau}^{i}_{m-2}-\tilde{\tau}^{k+\rho^{'}(i)}_{m-2})})\\
&( \prod^{m-3}_{p;\beta=0} \exp{-2 \pi i \tilde{\sigma}^{p}_{0} m_{\beta}})
\end{split}
\end{equation}
where $\eta=\sum^{m-3}_{\beta=1}\eta_{\beta}$.

Note that, $\sum_{{\rho_{\beta}}}\prod_{\beta=0,1,...,m-3}(-1)^{\rho_{\beta}}\prod_p \exp{-2 \pi i \tilde{\sigma}^{\rho^{-1}_{0}....\rho^{-1}_{m-4}(p)}_{0} \tilde{\tau}^{\rho_{m-3}(p)}_{m-2}}=\sum_{\tilde{\rho}}(-1)^{\tilde{\rho}}\prod_p \exp{-2 \pi i \tilde{\sigma}^{p}_{0} \tilde{\tau}^{\tilde{\rho}(p)}_{m-2}}$, where, $\tilde{\rho}=\rho_{m-3} \circ \rho_{m-4} \circ....\circ \rho_0$.

Also, note that $T_{NS5}$ only depends on $\rho^{'} \circ \tilde{\rho}^{-1}$, so that $\tilde{\rho}$ can be taken to be a trivial permutation of the indices (the integrand gets multiplied by a factor of $2k!$, which we will ignore).

Plugging this back in the expression for $\tilde{T}_{NS5}$, we have,
\begin{equation}
\begin{split}
\tilde{T}_{NS5}&=\int  \prod_{\beta=1,..,m-3}\prod_p \frac{1}{\cosh{\pi(\tilde{\sigma}^p_{0}+\sum^{\beta}_{m-3} \eta_{\beta})}}\\
&\times(\sum_{\rho^{'}}(-1)^{\rho^{'}}\prod_i \exp{2 \pi i \tilde{\tau}^i_{m-2}(\tau^{'i}-\tilde{\sigma}^i_0-\eta-\eta_3)} \exp\{{-2 \pi i \tilde{\tau}^{k+i}_{m-2}(\tau^{'\rho^{'-1}(i)}+\tilde{\sigma}^{k+i}_0+\eta+\eta_4)}\} \frac{\sinh{\pi \tau^{'i}}}{\cosh{\pi \tau^{'i}}})\\
&\times\prod_p \frac{\exp{2 \pi i (\tilde{\sigma}^{p}_{0} -\eta^p_0)\tilde{\tau}^{p}_0}}{\cosh{\pi \tilde{\sigma}^p_{0}}}(\prod^{m-3}_{p;\beta=0} \exp{-2 \pi i \tilde{\sigma}^{p}_{0} m_{\beta}})\\
& =\int  \prod_{\beta=1,..,m-3}\prod_p \frac{1}{\cosh{\pi (\tilde{\sigma}^p_{0}+\sum^{\beta}_{m-3} \eta_{\beta})}}\\
&\times(\sum_{\rho^{'}}(-1)^{\rho^{'}}\prod_i \delta(\tau^{'i}-\tilde{\sigma}^i_0-\eta-\eta_3) \delta(\tau^{'\rho^{'-1}(i)}+\tilde{\sigma}^{k+i}_0+\eta+\eta_4) \frac{\sinh{\pi \tau^{'i}}}{\cosh{\pi \tau^{'i}}})\\ 
&\times\prod_p \frac{\exp{2 \pi i (\tilde{\sigma}^{p}_{0}-\eta^p_0) \tilde{\tau}^{p}_0}}{\cosh{\pi \tilde{\sigma}^p_{0}}} 
(\prod^{m-3}_{p;\beta=0} \exp{-2 \pi i \tilde{\sigma}^{p}_{0} m_{\beta}})\\
&=\int  \prod_{\beta=1,..,m-3}\prod_p \frac{1}{\cosh{\pi (\tilde{\sigma}^p_{0}+\sum^{\beta}_{m-3} \eta_{\beta})}}
(\sum_{\rho^{'}}(-1)^{\rho^{'}} \delta(\tilde{\sigma}^{\rho^{'-1}(i)}+\tilde{\sigma}^{k+i}_0+2 \eta+\eta_3+\eta_4) \frac{\sinh{\pi (\tilde{\sigma}^{i}_0+\eta+\eta_3)}}{\cosh{\pi(\tilde{\sigma}^{i}_0+\eta+\eta_3)}})\\ 
&\times \prod_p \frac{\exp{2 \pi i (\tilde{\sigma}^{p}_{0}-\eta^p_0) \tilde{\tau}^{p}_0}}{\cosh{\pi \tilde{\sigma}^p_{0}}} 
(\prod^{m-3}_{p;\beta=0} \exp{-2 \pi i \tilde{\sigma}^{p}_{0} m_{\beta}})
\end{split}
\end{equation}
where we have integrated out $\tau_{m-2}$ and $\tau^{'}$ in the second and the final step respectively.

Therefore, the final expression for $\tilde{T}_{NS5}$ becomes,

\begin{equation}
\begin{split}
\tilde{T}_{NS5}&=\int  (\prod_{\beta=1,..,m-3}\prod_p \frac{\exp{2 \pi i (\tilde{\sigma}^{p}_{0}-\eta^p_0) \tilde{\tau}^{p}_0}}{\cosh{\pi (\tilde{\sigma}^p_{0}+\sum^{\beta}_{m-3} \eta_{\beta})}}) \frac{1}{\prod_p \cosh{\pi \tilde{\sigma}^p_{0}}}(\prod_i \frac{\sinh{2 \pi (\tilde{\sigma}^{i}_0+\eta+\eta_3)}}{\cosh^2{\pi (\tilde{\sigma}^{i}_0+\eta+\eta_3)}})\\
& \times(\sum_{\rho^{'}}(-1)^{\rho^{'}}\prod_i \delta(\tilde{\sigma}^{k+\rho^{'}(i)}_0+\tilde{\sigma}^i_0+2 \eta+\eta_3+\eta_4)) (\prod^{m-3}_{p;\beta=0} \exp{-2 \pi i \tilde{\sigma}^{p}_{0} m_{\beta}})
\end{split}
\end{equation}

The S-dual partition function of the A-model is then given as,
\begin{equation}
\begin{split}
\tilde{Z}_A&=\int  \prod_{\beta=1,..,m-3}\prod_p \frac{1}{\cosh{\pi (\tilde{\sigma}^p_{0}+\sum^{\beta}_{m-3} \eta_{\beta})}} \frac{1}{\prod_p \cosh{\pi \tilde{\sigma}^p_{0}}} (\prod_i \frac{\sinh{2 \pi (\tilde{\sigma}^{i}_0+\eta +\eta_3)}}{\cosh^2{\pi (\tilde{\sigma}^{i}_0+\eta +\eta_3)}})\\
&\times\prod^{\frac{n}{2}}_{a=1} \frac{\prod_{p < l}\sinh{\pi(\sigma^{'p}_{a}-\sigma^{'l}_{a})}\sinh{\pi(\sigma^{'p}_{a+1}-\sigma^{'l}_{a+1})}}{\prod_{p,l}\cosh{\pi(\sigma^{'p}_{a}-\sigma^{'l}_{a+1})}}
(\frac{\prod_i \sinh{\pi 2 \sigma^{'i}_{\frac{n}{2}+1}}}{\prod_i \cosh^2{\pi\sigma^{'i}_{\frac{n}{2}+1}}}) (\prod^{\frac{n}{2}}_{p;a=1}  \exp{2 \pi i m^{f}_a (\sigma^{'p}_a-\sigma^{'\rho^{-1}_{a+1}(p)}_{a+1})})\\
&\times(\sum_{\rho^{'}}(-1)^{\rho^{'}}\prod_i \delta(\tilde{\sigma}^{k+\rho^{'}(i)}_0+\tilde{\sigma}^i_0+2\eta+\eta_3+\eta_4)) (\sum_{\rho} (-1)^{\rho} \delta(\sigma^{'k+i}_{\frac{n}{2}+1}+\sigma^{'\rho^{-1}(i)}_{\frac{n}{2}+1}))\\
&\times (\sum_{\rho_1}(-1)^{\rho_1} \prod_p \exp{-2 \pi i \sigma^{'p}_1 \tilde{\tau}^{\rho_1(p)}_0} \exp{2 \pi i (\tilde{\sigma}^{p}_0 -\eta^p_0)\tilde{\tau}^{p}_0}))\prod^{m-3}_{p;\beta=0} \exp{-2 \pi i \tilde{\sigma}^{p}_{0} m_{\beta}}
\end{split}
\end{equation}
By relabeling the integration variables, one can reduce all of $\rho^{'},\rho,\rho_1$ to trivial permutation. Now, we define $\tilde{\sigma}^i_0+\eta+\eta_3 \rightarrow \tilde{\sigma}^i_0$ and $\tilde{\sigma}^{k+i}_0+\eta+\eta_4 \rightarrow \tilde{\sigma}^{k+i}_0$.

With this redefinition the delta functions imply,

\begin{equation}
\sigma^{'k+i}_{\frac{n}{2}+1}=-\sigma^{'i}_{\frac{n}{2}+1}
\end{equation}
\begin{equation}
\tilde{\sigma}^{k+i}_{0}=-\tilde{\sigma}^{i}_{0}
\end{equation}
Also, on integrating out $\tilde{\tau}^{p}_0$, the resultant delta function relates $\sigma^{'}_1$ and $\tilde{\sigma}_0$ as follows:

\begin{equation}
\sigma^{'i}_{1}=\tilde{\sigma}^{i}_{0}-\eta -\eta_1 -\eta_3
\end{equation}
\begin{equation}
\sigma^{'k+i}_{1}=\tilde{\sigma}^{k+i}_{0}-\eta -\eta_2 -\eta_4
\end{equation}
Now, integrating out $\sigma^{'k+i}_{\frac{n}{2}+1}$ and $\sigma^{'k+i}_{0}$,
\begin{equation}
\begin{split}
\tilde{Z}_A&=\int  \prod_{\beta=1,..,m-3}\prod_i \frac{1}{\cosh{\pi (\tilde{\sigma}^i_{0}-\eta -\eta_3+\sum^{\beta}_{m-3} \eta_{\beta})} \cosh{\pi (\tilde{\sigma}^i_{0}+\eta+\eta_3-\sum^{\beta}_{m-3} \eta_{\beta})}} \\
&\times \frac{1}{\cosh{\pi (\tilde{\sigma}^i_{0}-\eta -\eta_3)} \cosh{\pi (\tilde{\sigma}^i_{0}+\eta +\eta_4)}}\\
&\times(\prod_i \frac{\sinh{2 \pi (\tilde{\sigma}^{i}_0)}}{\cosh^2{\pi (\tilde{\sigma}^{i}_0)}}) \prod^{\frac{n}{2}}_{a=1} \frac{\prod_{p < l}\sinh{\pi(\sigma^{'p}_{a}-\sigma^{'l}_{a})}\sinh{\pi(\sigma^{'p}_{a+1}-\sigma^{'l}_{a+1})}}{\prod_{p,l}\cosh{\pi(\sigma^{'p}_{a}-\sigma^{'l}_{a+1})}}(\frac{\prod_i \sinh{\pi 2 \sigma^{'i}_{\frac{n}{2}+1}}}{\prod_i \cosh^2{\pi\sigma^{'i}_{\frac{n}{2}+1}}})\\
&\times (\prod^{\frac{n}{2}}_{p;a=1} \exp{2 \pi i m^{f}_a (\sigma^{'p}_a-\sigma^{\rho^{-1}_{a+1}(p)}_{a+1})})
\end{split}
\end{equation}

with $\sigma^{'i}_1$ given as a function of $\tilde{\sigma}^{i}_{0}$ as above. Note that the contribution of the term $\prod^{m-3}_{p;\beta=0} \exp{-2 \pi i \tilde{\sigma}^{p}_{0} m_{\beta}}$ drops out, since $\tilde{\sigma}^{k+i}_{0}=-\tilde{\sigma}^{i}_{0}$.

For $\frac{1}{\cosh{\pi (\tilde{\sigma}^i_{0}-\eta -\eta_3)} \cosh{\pi (\tilde{\sigma}^i_{0}+\eta +\eta_4)}}$ to be the contribution of a fundamental hyper of $Sp(k)$ in the mirror theory, we need $\eta_3=\eta_4$.

Also,

$\prod_{p < l}\sinh{\pi(\sigma^{'p}_{1}-\sigma^{'l}_{1})} =\prod_{i < j}\sinh^2{\pi(\tilde{\sigma}^{i}_{0}-\tilde{\sigma}^{j}_{0})}\prod_{i,j}\sinh{\pi(\tilde{\sigma}^{i}_{0}+\tilde{\sigma}^{j}_{0}+\eta_2+\eta_4-\eta_1-\eta_3)}$. \\

For the above to be a contribution to a $Sp(k)$ vector multiplet in the mirror theory, we need to set, $\eta_1=\eta_2$. Note that the partition function computation, therefore, reproduces one of the predictions from the Type IIB brane construction,namely, $\eta_1=\eta_2,\eta_3=\eta_4$.

Finally, the remaining phase factors can be rewritten as,\\
$\prod^{\frac{n}{2}}_{a=1} \prod_p \exp{2 \pi i m^{f}_a (\sigma^{'p}_a-\sigma^{'\rho^{-1}_{a+1}(p)}_{a+1})}=\prod^{\frac{n}{2}}_{a=2} \prod_p\exp{2 \pi i \sigma^{'p}_a (m^f_a-m^f_{a-1})}$. Thus,we can recast $\tilde{Z}_A$ as follows:

\begin{equation}
\begin{split}
\tilde{Z}_A&=\int  \prod^{m-3}_{\beta=1}\frac{1}{\prod_{i}\cosh{\pi(\tilde{\sigma}_0^i-\eta -\eta_3+\sum^{\beta}_{m-3} \eta_{\beta})}\cosh{\pi(\tilde{\sigma}_0^i+\eta+\eta_3-\sum^{\beta}_{m-3} \eta_{\beta})}}\\
&\times\frac{1}{\prod_{i}\cosh{\pi (\tilde{\sigma}^i_{0}-\eta -\eta_3)} \cosh{\pi (\tilde{\sigma}^i_{0}+\eta +\eta_3)} \cosh^2{\pi \tilde{\sigma}^i_{0}}} 
\frac{1}{\prod_{i}\cosh^2{\pi\sigma^{'i}_{\frac{n}{2}+1}}} \\
&\times F(\tilde{\sigma}_0,\sigma^{'}_2;-(\eta+\eta_1+\eta_3)) F(\sigma^{'}_{\frac{n}{2}+1},\sigma^{'}_{\frac{n}{2}};0)\\
&\times\frac{\prod_{p < l,\beta=2,..,\frac{n}{2}-1}\sinh{\pi(\sigma^{'p}_{\beta}-\sigma^{'l}_{\beta})}\sinh{\pi(\sigma^{'p}_{\beta+1}-\sigma^{'l}_{\beta+1})}}{\prod_{p,l,\beta=1,..,\frac{n}{2}-2}\cosh{\pi(\sigma^{'p}_{\beta}-\sigma^{'l}_{\beta+1})}}(\prod^{\frac{n}{2}}_{a=2} \prod_p\exp{2 \pi i \sigma^{'p}_a (m^f_a-m^f_{a-1})})
\end{split}
\end{equation}
Finally, relabeling, $\tilde{\sigma}^i_0 \rightarrow \sigma^i_1, \tilde{\sigma}^{'p}_{\beta} \rightarrow \tilde{\sigma}^{'p}_{\beta-1}$(for $\beta=2,3,...\frac{n}{2}$),$\sigma^{'i}_{\frac{n}{2}+1} \rightarrow \sigma^{i}_{\frac{n}{2}}$, we conclude that,

\begin{equation}
\tilde{Z}_A=Z_B
\end{equation}
with certain relations between the mass and FI parameters, which precisely state the mirror map in this case.

\noindent \textbf{Mirror Map}

Comparing the expressions for $\tilde{Z}_A$ and $Z_B$, we see that,

\begin{equation}
m_{bif}=-(\eta+\eta_1+\eta_3)
\end{equation}
All other bi-fundamental masses in the B-model are zero.

The $(m-2)$ fundamental masses in the B-model are given as,

\begin{equation}
M^f_a=-(\eta + \eta_3 - \sum^{a}_{m-3} \eta_{\beta})=-(\sum^{a-1}_{1} \eta_{\beta} + \eta_3);  a=1,2,...,m-3
\end{equation}
\begin{equation}
M^f_{m-2}=-(\eta + \eta_3),M^f_{m-1}=0,M^f_{m}=0
\end{equation}
The non-zero fundamental masses, together with the non-zero bi-fundamental mass gives a total of $(m-1)$ mass parameters as predicted by the Type IIB description.
On the other hand, the $n/2 -1$ FI parameters of the B-model are given as,

\begin{equation}
\tilde{\eta}_{\beta}=m^f_{\beta +1}-m^f_{\beta}, \beta=1,2,...,\frac{n}{2}-1
\end{equation}
Since only $\frac{n}{2}-1$ of the masses in the A-model are independent, as we argued before, we may set $m_1$ to zero, which slightly modifies the mirror map as follows:

\begin{equation}
\tilde{\eta}_{1}=m^f_2,\eta^{B}_{\beta}=m^f_{\beta +1}-m^f_{\beta}, \beta=2,...,\frac{n}{2}-1
\end{equation}

Therefore, to summarize, the total number of independent mass-parameters is $\frac{n}{2}-1$ and the number of independent FI parameters is $m-1$. In the B-model, the total number of independent (non-zero) mass parameters is $(m-2)+1=m-1$ and the number of FI parameters is $\frac{n}{2}-1$. Thus, there seems to be a perfect match of allowed deformations in the two theories, under mirror symmetry.\\

The above computation can be trivially modified to include fundamental hypers for the $U(2k)$ factors, as long as $U(k)_1$ and $U(k)_2$ (and similarly $U(k)_3$ and $U(k)_4$) have equal number of fundamental hypers.\\

We would like to end this section by briefly commenting on the number of FI parameters and masses of the A-model that enter the mirror map computed above .  As shown earlier, duality between the A-model and the B-model seems to require that, in the A-model, $\eta_1=\eta_2$ and $\eta_3=\eta_4$, so that there are only $m-1$ independent FI parameters as opposed to the expected number $m$ for a $D_m$ quiver (corresponding to every $U(1)$ in $U(k)^4 \times U(2k)^{m-3}/ U(1)_D$, $U(1)_D$ being the diagonal $U(1)$ subgroup of the gauge group). The fundamental masses of $U(k)_1$ and $U(k)_2$ are also required to be pairwise equal, so that there are only $n/2$ independent fundamental hyper masses. Finally, the mass of the $U(k)_1 \times U(2k)_1$ bifundamental hyper has to be equal to the mass of the $U(k)_2 \times U(2k)_1$ bifundamental hyper (same is true for the masses of the $U(k)_3 \times U(2k)_{m-3}$ and the $U(k)_4 \times U(2k)_{m-3}$ bifundamental hypers).\\
Now, the $D_m$ quiver has a $\mathbb{Z}_2$ outer automorphism symmetry which acts by exchanging the gauge groups $U(k)_1$ and $U(k)_2$ (as well as $U(k)_3$ and $U(k)_4$) . The reduced number of FI parameters and hypermultiplet masses (and, in particular, the relations connecting some of the FI parameters mentioned above) can be readily explained if we require the A-model to be even under this outer automorphism (understood as a discrete gauge symmetry) so that the $\mathbb{Z}_2$ odd operators of the theory are projected out. The action of this symmetry on the space of moduli of the theory, in particular, projects out linear combinations of FI parameters and hypermultiplet masses which are odd under the aforementioned transformation. The space of FI parameters of the gauged theory , even under this discrete symmetry, is then restricted to  $\eta_1=\eta_2$ and $\eta_3=\eta_4$. Exactly analogous argument holds for the hypermultiplet masses.\\
We, therefore, conclude that the supersymmetric gauge theory dual to the B-model (described in figure 11(b)) is a $D_m$ quiver, specified in figure 11(a), which is also even under the action of the $\mathbb{Z}_2$ outer automorphism as a discrete gauge symmetry.\\

\section{Conclusion}
In this note, we have discussed the M-theory description of mirror symmetry in three dimensions for a large class of $\mathcal{N}=4$ quiver gauge theories, involving an eleven-dimensional supergravity solution with the geometry $\mathcal{M}=\mathbb{R}^{2,1} \times ALE_1 \times ALE_2$ and a configuration of four-form G-fluxes over the eight-dimensional transverse manifold $ALE_1\times ALE_2$. In particular, we argued that a pair of quiver gauge theories, mirror dual to each other, can be described by the  M-theory backgrounds $\mathcal{M}_1=\mathbb{R}^{2,1} \times ALF_1 \times ALE_2$ and $\mathcal{M}_2=\mathbb{R}^{2,1} \times ALE_1 \times ALF_2$ respectively, for generic values of gauge couplings. In the IR limit, which is also the strong coupling limit for the gauge theories, the dual theories, in question, flow to the same M-theory background, $\mathcal{M}_{IR}=\mathbb{R}^{2,1} \times ALE_1 \times ALE_2$. 

This M-theory interpretation readily allows one to use the $A-D-E$ classification of $\text{ALE}$ spaces to derive a systematic catalogue of the discrete families of mirror symmetric quiver gauge theory pairs.  In addition to the well-known examples (figures \ref{fig1l}-\ref{fig3l}), the procedure leads to a set of new mirror duals (figures \ref{fig4l}-\ref{fig8l}), the field content of which are determined from the corresponding Type IIA set-up, as previously explained. For the $\mathbb{Z}_n\times \mathbb{D}_{m-2}$ theories ($n$ even), we have presented the Type IIB brane construction to illustrate the map between the two sets of integers $\{w_i\}$ and $\{v_j\}$ (which parametrize the distributions of fundamental hypermultiplets on the dual quivers) and discussed relevant features of mirror maps for this class of dual theories.

The duality can be tested beyond the moduli space arguments , by directly computing partition functions of mirror-dual theories on $S^3$ in the IR limit using localization techniques. An agreement of the partition functions essentially gives a proof of the duality for the pair of theories under consideration. After demonstrating the procedure in one of the simpler and more well-known example of mirror duals ($\mathbb{Z}_n\times \text{Trivial}$), we perform a computation for the $\mathbb{Z}_n\times \mathbb{D}_{m-2}$ ($n$ even) case . In addition to confirming the duality,the partition function computation provides a way to explicitly determine the mirror-map for this family of dual theories. We observe that the mirror map obtained in this case is in complete agreement with the predictions from the Type IIB set-up.

\section{Acknowledgements}
The author would like to thank Jacques Distler for numerous discussions that helped shape the project. The author benefitted greatly from discussions with Andy Neitzke
and Yuji Tachikawa at different stages of the work.

This work was supported by the National Science Foundation under Grant Number PHY-0969020.

\renewcommand{\theequation}{A.\arabic{equation}}
\setcounter{equation}{0}
\section*{Appendix A}
\addcontentsline{toc}{section}{Appendix A}

\subsection{Dual Theories for $\mathbb{Z}_n \times \mathbb{D}_{m-2}$ Singularity}

\textbf{Case 1: n even}

\noindent \textbf{A-Model} : The A-Model corresponds to the case where the M-theory circle lies along the ALF obtained by deforming $\frac{\mathbb{C}^2}{\mathbb{Z}_n}$ . The resulting Type IIA background consists of $n$ $D6$ branes wrapping the $\frac{\mathbb{C}^2}{\mathbb{D}_{m-2}}$. There are $k$ $D2$ branes parallel to the $D6$. The $D2$ branes wrap the directions $0,1,2$,while $D6$ wraps $0,1,2,3,4,5,6$.

Since this background does not involve orientifold planes, one only needs to compute the orbifold projection of the spectrum of open string states under $\Gamma=\mathbb{D}_{m-2}$ from the $D2-D2$ and the $D2-D6$ sectors to obtain the field content of the world-volume theory on the $D2$ branes. It is sufficient to deal with the NS-sector fields (bosons) only, since the R-sector fields (fermions) will naturally assemble themselves into appropriate supermultiplets because of the $\mathcal{N}=4$ supersymmetry of the background.

\noindent \textbf{D2-D2 Spectrum} The NS-sector fields, arising from D2-D2 open strings are: $A^{\mu}_{ij} (\mu=0,1,2)$,$X^{I}_{ij} (I=3,4,5,6)$,$Y^a_{ij} (a=7,8,9)$; where $i,j=1,2,....,N$ are the Chan-Paton factors with $N=k|\Gamma|$ being the dimension of the covering space. The indices $\mu$ and $a$ are invariant under the action of $\Gamma$. $X^I s$ can be arranged into a pair of complex fields on which $\Gamma$ acts by the usual $SU(2)$ action on $\mathbb{C}^2$. The action of the orbifold projection on the fields is given by,

\begin{equation}
A_{\mu}=\gamma(g) A_{\mu} \gamma(g)^{-1}
\end{equation}
\begin{equation}
X^I=R^I_J \gamma(g) X^J \gamma(g)^{-1}
\end{equation}
with the $Y^a$s transforming the same way as the $A_{\mu}$s.

The gauge field and matter content can be obtained by directly solving the above equations. However, the field content can also be derived from a more algebraic approach by counting the $\Gamma$-invariant homomorphisms, noting that $A^{\mu}_{ij},Y^m_{ij} \in Hom[\mathbb{C}^N,\mathbb{C}^N]$ and $X^I_{ij} \in 2  \otimes Hom[\mathbb{C}^N,\mathbb{C}^N]$.

For the gauge field,

\begin{equation}
(A^{\mu}_{ij})_{\Gamma}=Hom[\mathbb{C}^N,\mathbb{C}^N]_{\Gamma}=[Hom(\oplus_i \mathbb{C}^{kn_i}\otimes r_i,\oplus_j \mathbb{C}^{kn_j}\otimes r_j)]_{\Gamma}=\oplus_{i,j}\mathbb{C}^{kn_i *}\otimes \mathbb{C}^{kn_j}\otimes [r_i^* \otimes r_j]_{\Gamma}
\end{equation}
\begin{equation}
=\oplus_i \mathbb{C}^{kn_i *} \otimes \mathbb{C}^{kn_i}=\oplus_i \mathbb{C}^{k dimr_i *} \otimes \mathbb{C}^{kdimr_i}
\end{equation}
\begin{equation}
=\oplus_i Hom(\mathbb{C}^{k dimr_i},\mathbb{C}^{k dimr_i})
\end{equation}
Here, $(A^{\mu}_{ij})_{\Gamma}$ denotes the $\Gamma$-invariant homomorphism. In the second line, we have used the fact that $[r_i^* \otimes r_j]_{\Gamma}=\delta_{ij}$ (by Schur'{}s Lemma) where $r_i,r_j$ are irreducible representations of $\Gamma=\mathbb{D}_{m-2}$ (the notation used above is a bit sloppy since it uses the same symbol for a representation and the vector space it is defined on).

We also assume that the representation of the orbifold group is regular i.e. the irrep $r_i$ appears $dimr_i$ number of times in the decomposition of the N-dimensional representation into irreps of $\mathbb{D}_{m-2}$. The sum in the last step is over all the irreps of $\mathbb{D}_{m-2}$ - 4 of them 1-dimensional and $(m-3)$ of them 2-dimensional. The gauge group can be immediately read off from above:

\begin{equation}
G_{gauge}= U(k)^4 \times U(2k)^{m-3}
\end{equation}
The analysis is identical for the $Y^a$ fields - orbifold projection yields 3 scalars (one for each $a$) in the adjoint of every factor in the gauge group. Thus the orbifold projected $A_{\mu}$ and $Y_a$,taken together, give the bosonic parts of 4 $U(k)$ vector multiplets and $(m-3)$ $U(2k)$ vector multiplets. For the $X^I$, we similarly have,

\begin{equation}
[2 \otimes Hom(\mathbb{C}^N,\mathbb{C}^N)]_{\Gamma}=\oplus_{i,j}a^{(2)}_{ij} Hom(\mathbb{C}^{k dimr_i},\mathbb{C}^{k dimr_j})
\end{equation}
where $a^{(2)}_{ij}r_j=2 \otimes r_i$ and $a^{(2)}_{ij}$ can be directly read off from the extended Dynkin diagram corresponding to  $\mathbb{D}_{m-2}$. The 4 outer nodes in the Dynkin diagram (see figure \ref{fig4l} (a)) represent the 4 $U(k)$ groups and the remaining nodes represent the $U(2k)$ groups. Then $a^2_{ij}$ is given by the number of lines connecting the $i$th and the $j$th node. So the orbifold projected $X^I$ fields assemble into the bosonic parts of hypermultiplets (4 real scalars) in the bi-fundamental of $U(k_i) \times U(k_j)$ for each pair of nodes $(i,j)$ for which $a^2_{ij} \neq 0$.

\noindent \textbf{D2-D6 Spectrum} For a single pair of D2-D6 branes, the DN/ND open string ground states yield a $D=3,\mathcal{N}=4$ hypermultiplet, whose scalars are given by the complex doublet $h^A$ ($\tilde{h}_A$ for ND).The two doublets are related by complex conjugation,

$\epsilon_{AB}(h^B)^{*}=\tilde{h}^A$

In case of multiple branes, the CP structure of the scalars is $h^A_{iI}$, where $i=1,2,...,N$ as before and $I=1,2,...,n$ represent the CP factors corresponding to the $D6$ branes. The orbifold action on these fields is given by,

\begin{equation}
h^A_{iI}=\gamma_{ij}(g) h^A_{jJ} \gamma^{-1}_{JI}(g)
\end{equation}
where, $\gamma_{JI}(g)$ gives the action of the $\mathbb{D}_{m-2}$ orbifolding on the CP factors of the D6 branes.

We may choose the orbifold action to act trivially on the $D6$ branes. In that case, the orbifold projected $h^A$ fields, give the bosonic part of $n$ hypermultiplets in the fundamental of one of the $U(k)$s.

For a general non-trivial action on the $D6$ CP factors, we note that, $h_{iI}\in Hom(\mathbb{C}^N,\mathbb{C}^n)$. Now, as before,

\begin{equation}
Hom[\mathbb{C}^N,\mathbb{C}^n]_{\Gamma}=[Hom(\oplus_i \mathbb{C}^{kn_i}\otimes r_i,\oplus_j \mathbb{C}^{w_j}\otimes r_j)]_{\Gamma}=\oplus_{i,j}\mathbb{C}^{kn_i *}\otimes \mathbb{C}^{w_j}\otimes [r_i^* \otimes r_j]_{\Gamma}
\end{equation}
\begin{equation}
=\oplus_i \mathbb{C}^{kn_i *} \otimes \mathbb{C}^{w_i}=\oplus_i \mathbb{C}^{k dimr_i *} \otimes \mathbb{C}^{w_i}
\end{equation}
\begin{equation}
=\oplus_i Hom(\mathbb{C}^{k dimr_i},\mathbb{C}^{w_i})
\end{equation}
so that the $i$th factor in the gauge group has $w_i$ fundamental hyper, subject to the condition, $\sum_i {w_i dimr_i}=n$

The particle content of the A-model can then be summarized as in the quiver diagram figure \ref{fig4l} (a).

Given the particle content, the dimensions of the Coulomb and the Higgs branch are as follows:

\begin{equation}
dim M^A_C=2k(m-1)
\end{equation}
\begin{equation}
dim M^A_H=nk
\end{equation}
\noindent \textbf{B-Model} : The B-Model corresponds to the case where the M-theory circle lies along the ALF obtained by deforming  $\frac{\mathbb{C}^2}{\mathbb{D}_{m-2}}$. The Type IIA background now consists of $m$ $D6$ branes near an $O6$ plane, wrapping the $\frac{\mathbb{C}^2}{\mathbb{Z}_n}$. There are again $k$ $D2$ branes parallel to the $D6$. The $D2$ branes wrap the directions $0,1,2$,while $D6$ wraps $0,1,2,3,4,5,6$.

\noindent \textbf{D2-D2 Spectrum} The NS sector fields are now projected under the action of both the orbifold and the orientifold group. The orbifold action on the fields are precisely the same (albeit under a different orbifold group,viz. $\mathbb{Z}_n$) and the orientifold action is given as follows:

\begin{equation}
A_{\mu}=-\gamma(\Omega \mathbb{Z}_2) A_{\mu}^T \gamma(\Omega \mathbb{Z}_2)^{-1}  \label{gor1}
\end{equation}
\begin{equation}
X_{I}=\gamma(\Omega \mathbb{Z}_2) X_{I}^T \gamma(\Omega \mathbb{Z}_2)^{-1}
\end{equation}
\begin{equation}
Y_{a}=-\gamma(\Omega \mathbb{Z}_2) Y_{a}^T \gamma(\Omega \mathbb{Z}_2)^{-1}
\end{equation}
The matrix $\gamma(\Omega \mathbb{Z}_2)$ has to satisfy the following consistency conditions:

\begin{equation}
(\Omega \mathbb{Z}_2)^2=\mathbb{I}:\gamma(\Omega \mathbb{Z}_2)=-\gamma(\Omega \mathbb{Z}_2)^T    \label{con1}
\end{equation}
\begin{equation}
\Omega g=g \Omega : \gamma(g) \gamma(\Omega \mathbb{Z}_2) \gamma(g)^T=\gamma(\Omega \mathbb{Z}_2)   \label{con2}
\end{equation}
where $g \in \mathbb{Z}_n$ is the generator of the orbifold group. We choose the following regular representation $\gamma (g)$ on the CP factors:

\begin{equation}
\gamma(g)= \left(\begin{matrix}
             I_{2k} & . & . &. & .\\
             . & \xi I_{2k} &. &.& .\\
             . & .          &. &.&.\\
             . & .          &. &.&.\\
             . & .          &. &.&\xi^{n-1} I_{2k}\\  
             \end{matrix}\right)
\end{equation}
where $I_{2k}$ is a $2k \times 2k$ identity matrix. Let $\gamma(\Omega\mathbb{Z}_2)=A_{i,\alpha;j,\beta}$, where $A_{ij}$ is a $(2k \times 2k)$ block and $i,j=1,2,....,n$. Then, equation \ref{con2} implies:

\begin{equation}
\xi^{i+j} A_{i,\alpha;j,\beta}=A_{i,\alpha;j,\beta}
\end{equation}
Hence,$i+j=n$ for $A_{i,\alpha;j,\beta} \neq 0$. On the other hand,equation \ref{con1} implies:

\begin{equation}
A_{i,\alpha;n-i,\beta}=-A_{n-i,\beta;i,\alpha}
\end{equation}
with $i=0$ being identified with $i=n$. Thus, choosing $A_{i,\alpha;n-i,\beta}=\delta_{\alpha\beta}$, for $0 < i < \frac{n}{2}$, we have $A_{n-i,\beta;i,\alpha}=-\delta_{\alpha\beta}$. For $i=0,\frac{n}{2}$, we have $A_{i,\alpha;i,\beta}=-A_{i,\alpha;i,\beta}^T=\epsilon_{2k}$, where $\epsilon_{2k}$ is the completely antisymmetric matrix with non-zero entries $\pm 1$. Thus, taken together the matrix $\gamma(\Omega\mathbb{Z}_2)$ is given as:

\begin{equation}
\gamma(\Omega\mathbb{Z}_2)= \left(\begin{matrix}
             \epsilon_{2k} & 0 & . &. & .& .& 0\\
             0&0 & 0 &. &.& .&0& I_{2k}\\
             0&0 & 0 &. &.& 0 &.&0\\
             0&0& 0 &. &0&I_{2k}&0& 0\\
             0&0& 0 &0&\epsilon_{2k}&0&.& 0\\ 
             0&0&0 &-I_{2k} &0&.&0& 0\\
             0&0&-I_{2k} &. &0&.&0& 0\\
             0&-I_{2k}&. &. &0&.&0& 0\\
             \end{matrix}\right)
\end{equation}
Now, the $\mathbb{Z}_n$ orbifolding reduces the gauge group to $U(2k)^n$, so that the matrix $A_{\mu}$ can be expressed as $A_{\mu}=diag[A_0,A_1,A_2,....,A_{n-2},A_{n-1}]$, where each $A_i$ is a $2k \times 2k$ matrix. From equation \ref{gor1}, we have:

\begin{equation}
A_i=-A_{n-i}^T, 0 < i < \frac{n}{2}
\end{equation}
\begin{equation}
A_0=-\epsilon_{2k} A_{0}^T \epsilon_{2k}^{-1}
\end{equation}
\begin{equation}
A_{\frac{n}{2}}=-\epsilon_{2k} A_{\frac{n}{2}}^T \epsilon_{2k}^{-1}
\end{equation}
Thus, the independent blocks consist of $A_0$ and $A_{\frac{n}{2}}$ which are in the adjoint of $Sp(k)$ and $A_1,A_2,.....,A_{\frac{n}{2}-1}$ which are in the adjoint of $U(2k)$s. The gauge group, on orientifold projection, is given by:

\begin{equation}
G_{gauge}= Sp(k) \times U(2k)^{\frac{n}{2}-1} \times Sp(k)
\end{equation}
The $Y^a$s obey the same projection equation under orientifolding, giving three scalars in the adjoint of each of the component groups. Together with the gauge field, they constitute the bosonic part of the $\mathcal{N}=4$ vector multiplet for each of the component groups.

As mentioned before, we assemble the $X_I$s into the complex doublet $X^A$ ($A=1,2$), diagonalizing the $SU(2)$ action.

The $\mathbb{Z}_n$ orbifolding implies,$\xi^{i-j\pm 1} X_{i,\alpha;j,\beta}=X_{i,\alpha;j,\beta}$, so that the non-zero blocks are $X_{i,i+1} \in Hom(\mathbb{C}^{2k},\mathbb{C}^{2k})$, with $i=0$ being identified with $i=n$ as before. The orientifold projection implies,

\begin{equation}
X_{i,i+1}=X_{n-i-1,n-i}^T, 0 < i < \frac{n}{2}
\end{equation}
\begin{equation}
X_{0,1}=\epsilon_{2k} (X_{n-1,0})^T
\end{equation}
\begin{equation}
X_{\frac{n}{2},\frac{n}{2}+1}=-\epsilon_{2k} X_{\frac{n}{2}-1,\frac{n}{2}}^T
\end{equation}
$X_{i,i+1}$ gives the bosonic part of a hypermultiplet in the bi-fundamental of $U(2k)_i \times U(2k)_{i+1}$ with $i=1,2,...,\frac{n}{2}-2$, while $X_{0,1}$ and $X_{\frac{n}{2}-1,\frac{n}{2}}$ are in the bi-fundamental of $Sp(k) \times U(2k)_1$ and $U(2k)_{\frac{n}{2}-1} \times Sp(k)$.

\noindent \textbf{D2-D6 Spectrum} Now consider the contribution of the $D2-D6$ open strings. In general, the scalars $h^A$s will be projected under both the $\mathbb{Z}_n$ orbifolding as well as the orientifolding.Under the orbifolding, $h^A \in Hom[\mathbb{C}^N,\mathbb{C}^{2m}]$, decomposes as, $Hom[\mathbb{C}^N,\mathbb{C}^{2m}]=\oplus_i Hom[\mathbb{C}^{2k},\mathbb{C}^{2v_i}]$ , where $\sum_i v_i=m$, the sum extends over the one-dimensional irreps of $\mathbb{Z}_n$, with not all $v_i$s being zero.\\
The $\mathbb{Z}_n$ projected scalars $h^A_{iI}$ (where $D2$ CP factor $i=1,2,...,2k|\Gamma|$ on orbifold projection and $D6$ CP factor $I=1,2,...,2m$) will now be projected under orientifolding. The orientifold action on the DN scalars is given by:
\begin{equation}
h^{2 *}_{i I}= -i \gamma_2(\Omega \mathbb{Z}_2)_{ii^{'}} h^1_{i^{'}I^{'}}\gamma_6(\Omega \mathbb{Z}_2)_{I^{'}I}
\end{equation}
\begin{equation}
h^{1 *}_{i I}= i \gamma_2(\Omega \mathbb{Z}_2)_{ii^{'}} h^2_{i^{'}I^{'}}\gamma_6(\Omega \mathbb{Z}_2)_{I^{'}I}
\end{equation}
The orientifold action gives a total of $8km$ real scalars, which can be assembled into the bosonic parts of $m$ fundamental hypermultiplets distributed among the factors of the gauge group, i.e. if the $i$th gauge group has $v_i$ fundamental hypers, then $\sum_i v_i=m$.

The particle content for the B-model can thus be summarized as in figure \ref{fig4l}(b) .

The dimension of the Coulomb branch and the Higgs branch for the B-model are as follows:
\begin{equation}
dim M^B_C=nk
\end{equation}
\begin{equation}
dim M^B_H=2k(m-1)
\end{equation}
Now, comparing with the results for the $A-model$, we have $dim M^B_C=dim M^A_H$ and $dim M^B_H=dim M^A_C$, as expected for mirror dual theories.\\

\noindent \textbf{Case 2: n odd}

\noindent \textbf{A-model}: The computation for the A-model is almost identical to the case where n is even. The D2-D2 spectrum is exactly the same - the $\mathbb{D}_{m-2}$ quiver gauge theory with gauge group $U(k)^2 \times U(2k)^{\frac{n}{2}-1} \times U(k)^2$ and bi-fundamental hypermultiplets. The D2-D6 spectrum gives $w_i$ hypermultiplets in the fundamental of the $i$th factor of the gauge group, such that $\sum_i w_i dimr_i=n$, where the sum extends over all the irreps of $\mathbb{D}_{m-2}$. Since n is odd, the above equation basically implies that the total number of hypers in the fundamental of the $U(k)$s have to be odd. This number is even for the case when n is even. The particle content of the A-model can then be summarized as in figure \ref{fig5l} (a). So the dimensions of the Coulomb and Higgs branches are:

\begin{equation}
dimM^A_C= 2k(m-1)
\end{equation}
\begin{equation}
dimM^A_H= nk
\end{equation}
\noindent \textbf{B-model}: The mirror theory,however, will have non-trivial differences with the n even case, essentially because the orientifolding matrix is different when n is odd.

\noindent \textbf{D2-D2 spectrum}: The NS sector fields are projected by the same orbifolding matrix. The orientifold matrix satisfying the consistency conditions \ref{con1}, \ref{con2} is given as:

\begin{equation}
\gamma(\Omega\mathbb{Z}_2)= \left(\begin{matrix}
             \epsilon_{2k} & 0 & . &. & .& .&0\\
             0&0 & 0 &. & .&0& I_{2k}\\
             0&0 & 0 &. & 0 &.&0\\
             0&0& . &0 &I_{2k}&0& 0\\ 
             0&0&0 &-I_{2k} &.&0& 0\\
             0&0&. &0 &.&0& 0\\
             0&-I_{2k}&0 &. &.&0& 0\\
             \end{matrix}\right)
\end{equation}
Proceeding in the same way as before, the orientifold-projected gauge fields and the scalar fields $Y_a$s assemble into the bosonic parts of vector multiplets for the gauge group,

\begin{equation}
G_{gauge}= Sp(k)\times U(2k)^{\frac{n-1}{2}}
\end{equation}
The orientifold-projected $X^I$s form the bosonic parts of hypers in the bi-fundamental of $Sp(k)\times U(2k)_1$ and $U(2k)_i\times U(2k)_{i+1}$ (for $1< i < \frac{n-3}{2}$). $U(2k)_{\frac{n-1}{2}}$ has one hypermultiplet in the anti-symmetric representation.

\noindent \textbf{D2-D6 spectrum}: The action of orbifolding on the DN scalars is precisely the same as in the n even case. On orientifolding, we have a total of $8km$ real scalars which assemble into hypermultiplets in the fundamental of the different factors in the gauge group, so that if $v_i$ is the number of fundamental hypers for the $i$th factor in the gauge group, then $\sum_i v_i=m$.

The particle-content of the B-model in the n odd case is summarized in figure \ref{fig5l} (b). The dimensions for the Coulomb and Higgs branches are:

\begin{equation}
dimM^B_C= k+\frac{n-1}{2}(2k)=nk
\end{equation}
\begin{equation}
dimM^B_H= 2km+k(2k-1)+\frac{n-1}{2}(4k^2)-(2k^2+k)-\frac{n-1}{2}(4k^2)=2k(m-1)
\end{equation}
Comparing with the results for the A-model, we see $dimM^A_C=dimM^B_H$ and $dimM^A_H=dimM^B_C$ - as expected of mirror duals.

\subsection{Dual Theories for $\mathbb{D}_{n-2} \times \mathbb{D}_{m-2}$ Singularity}

We consider the case of even m and even n in detail. The computation for odd m (or n) is almost identical - the difference comes from the fact that the one-dimensional representations of a binary dihedral group of odd order are different from those of even order.

\noindent \textbf{A-Model} The A-Model corresponds to choosing the M-theory circle to lie along the ALF obtained by deforming $\frac{\mathbb{C}^2}{\mathbb{D}_{n-2}}$ . In the Type IIA picture, we have $n$ $D6$ branes near an $O6$ plane, wrapping the $\frac{\mathbb{C}^2}{\mathbb{D}_{m-2}}$, with $k$ $D2$ branes parallel to the $D6$.

\noindent \textbf{D2-D2 Spectrum} The NS-sector fields $A^{\mu}$,$X^I$,$Y^a$ have CP factors $i=1,2,..., 2k|\Gamma|$, where $\Gamma=4(m-2)$ is the total number of elements of $\mathbb{D}_{m-2}$. First we consider in a little more detail the $\mathbb{D}_{m-2}$ orbifolding which acts on the CP factors by the usual regular representation. The fields need to be invariant under the action of the two generators of the $\mathbb{D}_{m-2}$ group, $a$ and $b$ and in the regular representation these are as follows:

\begin{equation}
\gamma(a)= \left(\begin{matrix}
             I_{2k} & 0 & . &. & .& .& .&0\\
             0&-I_{2k} & 0 &. &.& .&0& 0\\
             0&0 & I_{2k} &. &.& 0 &.&0\\
             0&.& . &-I_{2k} &0&.&.& 0\\
             0&.& . &0&\tilde{a}_1&0&.& 0\\ 
             0&.&. &0 &0&.&0& 0\\
             0&.&. &. &.&0&.& 0\\
             0&.&. &. &0&.&0& \tilde{a}_{m-3}\\
             \end{matrix}\right)
\end{equation}
\begin{equation}
\gamma(b)= \left(\begin{matrix}
             I_{2k} & 0 & . &. & .& .& .&0\\
             0&I_{2k} & 0 &. &.& .&0& 0\\
             0&0 &-I_{2k} &. &.& 0 &.&0\\
             0&.& . &-I_{2k} &0&.&.& 0\\
             0&.& . &0&\tilde{b}_1&0&.& 0\\ 
             0&.&. &0 &0&.&0& 0\\
             0&.&. &. &.&0&.& 0\\
             0&.&. &. &0&.&0& \tilde{b}_{m-3}\\
             \end{matrix}\right)
\end{equation}
where

$\tilde{a}_l=\left(\begin{matrix}
                     \exp{(i\pi l/(m-2))}I_{2k}&0&0&0\\
                     0&\exp{-(i\pi l/(m-2))}I_{2k}&0&0\\
                     0&0&\exp{(i\pi l/(m-2))}I_{2k}&0\\
                     0&0&0&\exp{-(i\pi l/(m-2))}I_{2k}\\
                    \end{matrix}\right)$ and

$\tilde{b}_l=\left(\begin{matrix}
                     0&0&0&i^l I_{2k}\\
                     0&0&i^l I_{2k}&0\\
                     0&i^l I_{2k}&0&0\\
                     i^l I_{2k}&0&0&0\\
                    \end{matrix}\right)$

with $l=1,2,...,m-3$. The orientifold action $\gamma(\Omega \mathbb{Z}_2)$ has to obey consistency conditions w.r.t both the generators, i.e.,

\begin{equation}
\gamma(a) \gamma(\Omega \mathbb{Z}_2) \gamma(a)^T=\gamma(\Omega \mathbb{Z}_2)
\end{equation}
\begin{equation}
\gamma(b) \gamma(\Omega \mathbb{Z}_2) \gamma(b)^T=\gamma(\Omega \mathbb{Z}_2)
\end{equation}
Denoting the orientifold matrix as $\gamma(\Omega \mathbb{Z}_2)_{M,\alpha;N,\beta}$, where $M,N$ label the different irreps in the regular representation, including their multiplicities and $\alpha,\beta$ run from $1$ to $2k$. $M,N=i,j$ denote the 1-dimensional irreps while $M,N=I,J$ denote the 2-dimensional irreps. With this notation, the solution for $\gamma(\Omega \mathbb{Z}_2)$ is given as follows:

\begin{equation}
\gamma(\Omega \mathbb{Z}_2)_{i;j}=0, i \neq j
\end{equation}
\begin{equation}
\gamma(\Omega \mathbb{Z}_2)_{i;j}=\left(\begin{matrix}
                     0&I_{k}\\
                     -I_{k}&0\\
                    \end{matrix}\right), i = j
\end{equation}
\begin{equation}
\gamma(\Omega \mathbb{Z}_2)_{I;J}=0, I \neq J
\end{equation}
\begin{equation}
\gamma(\Omega \mathbb{Z}_2)_{I;J}=\left(\begin{matrix}
                     0&I_{2k}&0&0\\
                     -I_{2k}&0&0&0\\
                     0&0&0&-I_{2k}\\
                     0&0&I_{2k}&0\\
                    \end{matrix}\right), I= J (even)
\end{equation}
\begin{equation}
\gamma(\Omega \mathbb{Z}_2)_{I;J}=\left(\begin{matrix}
                     0&I_{2k}&0&0\\
                     -I_{2k}&0&0&0\\
                     0&0&0&I_{2k}\\
                     0&0&-I_{2k}&0\\
                    \end{matrix}\right), I= J (odd)
\end{equation}
and finally,

\begin{equation}
\gamma(\Omega \mathbb{Z}_2)_{i;I}=0.
\end{equation}
As before, the $D2-D2$ spectrum is obtained by orbifold-projecting the fields and orientifold-projecting with the form of $\gamma(\Omega \mathbb{Z}_2)$ as given above. The gauge-fields,for example, satisfy,

\begin{equation}
A^{\mu}=\gamma(a) A^{\mu} \gamma(a)^{-1}= \gamma(b) A^{\mu} \gamma(b)^{-1}=-\gamma(\Omega \mathbb{Z}_2) A^{\mu T} \gamma(\Omega \mathbb{Z}_2)^{-1}
\end{equation}
The resultant gauge group, is given by,

\begin{equation}
G_{gauge}= Sp(k)^4 \times SO(4k)^{\frac{m}{2}-1} \times Sp(2k)^{\frac{m}{2}-2}
\end{equation}
The $X^I$s on projection, yield half-hypers in the bifundamental of $Sp(k) \times SO(4k)$ and half-hypers in the bifundamental of $Sp(2k) \times SO(4k)$ (see figure \ref{fig6l} (a) ).The $Y^a$s, as before, are projected in the same way as the gauge fields.

\noindent \textbf{D2-D6 Spectrum} Since the action of the orbifold on the $D6$ brane is chosen to be trivial, one can simply choose $\gamma_6(\Omega \mathbb{Z}_2)=I_{2n}$ ( $\gamma_6(\Omega \mathbb{Z}_2)$ has to be symmetric since $\gamma_2(\Omega \mathbb{Z}_2)$ is antisymmetric). The orbifolded fields, on orientifolding, give $n$ hypers (and not half-hypers) in the fundamental of one of the four $Sp(k)$s. A non-trivial choice of the orbifolding action on the $D6$ branes will lead to a the $i$th gauge group having $w_i$ fundamental hypers, such that $\sum_i w_i dimr_i=n$, where $r_i$ denotes the irrep of $\mathbb{D}_{m-2}$ corresponding to the $i$th node in the quiver. The field content of the theory can, therefore, be  summarized in figure \ref{fig6l} (a).

Given the particle content, the dimensions of the Coulomb and the Higgs branch are as follows:

\begin{equation}
dim M^A_C=2k(m-1)
\end{equation}
\begin{equation}
dim M^A_H=2k(n-1)
\end{equation}
\textbf{B-Model} The B-Model is simply obtained by exchanging $m$ and $n$ in quiver diagram for the A-Model and is summarized in figure \ref{fig6l} (b) .

The dimensions of the Coulomb and the Higgs branch,in this case, are:

\begin{equation}
dim M^B_C=2k(n-1)
\end{equation}
\begin{equation}
dim M^B_H=2k(m-1)
\end{equation}
Thus, $dim M^A_C=dim M^B_H$ and $dim M^A_H=dim M^B_C$, as expected for mirror dual theories.

\end{document}